# A Survey on Semantic Communications for Intelligent Wireless Networks

**Sridhar Iyer**
**Department of ECE, KLE Dr. M.S. Sheshgiri College of Engineering and Technology, Belagavi, Karnataka, India (e-mail:** sridhariyer1983@klescet.ac.in)

**Rajashri Khanai**
**Department of ECE, KLE Dr. M.S. Sheshgiri College of Engineering and Technology, Belagavi, Karnataka, India**

**Dattaprasad Torse**
**Department of CSE, KLE Dr. M.S. Sheshgiri College of Engineering and Technology, Belagavi, Karnataka, India**

**Rahul Jashvantbhai Pandya**
**Department of EE, Indian Institute of Technology-Dharwad, WALMI Campus, PB Road, Karnataka, India.**

**Khaled Rabie**
**Department of Engineering, Manchester Metropolitan University, Manchester, UK.**

**Krishna Pai**
**Department of ECE, KLE Dr. M.S. Sheshgiri College of Engineering and Technology, Belagavi, Karnataka, India**

**Wali Ullah Khan**
**Interdisciplinary Centre for Security, Reliability and Trust (SnT), University of Luxembourg, 1855 Luxembourg City, Luxembourg.**

**Zubair Fadlullah**
**Department of Computer Science, ATAC Building, Room AT5024, Lakehead University, ON P7B 5E1 Canada.**

## Abstract

With the upsurge in next-generation intelligent applications and exponentially increasing heterogeneous data traffic both, the industry and the research community have already begun conceptualizing the 6G technology. With deployment of the 6G technology, it is envisioned that the competitive edge of wireless technology will be sustained and the next decade's communication requirements will be stratified. Also, the 6G technology will aim to aid the development of a human society which is ubiquitous, intelligent, and mobile, simultaneously providing solutions to key challenges such as, coverage, capacity, providing intelligence, computing, etc. In addition, 6G technology will focus on providing the intelligent use-cases and applications using higher data-rates over the mill-meter (mm) waves and Tera-Hertz (THz) frequency. However, at higher frequencies multiple non-desired phenomena such as atmospheric absorption, blocking, etc., occur which create a bottleneck owing to resource (spectrum and energy) scarcity. Hence, following the same trend of making efforts towards reproducing at the receiver, the exact information which was sent by the transmitter, will result in a never ending need for higher bandwidth. A possible solution to such a challenge lies in semantic communications which focuses on the meaning (relevance/context) of the received data as opposed to only reproducing the correct transmitted data. This in turn will require less bandwidth, and will reduce the bottleneck due to the various undesired phenomenon. In this respect, the current article presents a detailed survey on the recent technological trends in regard to semantic communications for intelligent wireless networks. We focus on the semantic communications architecture including the model, and source and channel coding. Next, we detail the cross-layer interaction, and the various goal-oriented communication applications. We also present the overall semantic communications trends in detail, and identify the challenges which need timely solutions before the practical implementation of semantic communications within the 6G wireless technology. Our survey article is an attempt to significantly contribute towards initiating future research directions in the area of semantic communications for the next-generation intelligent 6G wireless networks.

# I. Introduction

With the human society advancing towards complete automation and remote management, it is only a matter of time before very high-capacity systems ensuring reliability and cost/power-efficiency become the need of the hour. Further, the exponential increase in traffic volume can be associated with the fast-paced development of next-generation technologies catering to the needs of emerging applications, such as augmented reality (AR) and virtual reality (VR), Internet of Everything (IoE), artificial intelligence (AI), machine learning (ML), robotics and automation, etc. In this regard, the 5G technology, which is currently being deployed globally, has been envisioned to provide the necessary advanced solutions [1]. However, a major drawback of the 5G technology is that it does not standardize the convergence of the various functions such as communication, sensing, security, computing, etc. [2]. With automated and intelligent service requirements, the various 5G functionalities will need to converge for provisioning the IoE scenario, and with a fast-paced move of the society towards a fully automated and intelligent network scenario, the 5G technology will also need to be upgraded in view of provisioning enhanced services with fully immersive experiences [3]. These factors point towards the requirement of high bandwidth and intelligence within the successor of the 5G technology viz., the 6G technology [4]. 6G-enabled wireless network will support the advanced services such as, massive Machine Type Communications (mMTC), enhanced Mobile BroadBand (eMBB), Ultra-Reliable, Low Latency Computation, Communication and Control (URLLCCC), etc. [5, 6]. These services will be accompanied by increased traffic volume and amount of devices which are connected in turn requiring high amount of bandwidth in turn posing immense challenges to accomplish the enhanced spatial- and spectral-efficiency. Also, the next-generation applications, such as holographic videos, VR, and ubiquitous connectivity, will require very high bandwidth, which cannot be supported by the mm-wave spectrum (30–100 GHz) used in the 5G technology. Therefore, to obtain a broader radio spectrum bandwidth and extremely wideband channels with tens of GHz-wide bandwidth, Tera-Hertz (THz) communication (i.e., 0.1-10 THz) looks a promising option [7].

From the above it can be easily inferred that, for providing solutions to the most advanced applications, every subsequent wireless network generation moves the operation to higher frequency(s). This is mainly the result of an effort to reproduce the exact data (bits/symbols) at the receiver which is sent by the transmitter, and is referred to as the 'Level 1 (technical) problem' in the article by Shannon [8]. In fact, over the past decades, solutions to the Level 1 problem has resulted in multiple significant advancements such as, multiple-input multiple-output (MIMO) technology, novel designs of waveform, multi-user interference minimization, network function virtualization (NFV), software defined networking (SDN), network slicing, etc. With current trend of next-generation applications requiring intelligent solutions, it is observed that there is a similar inclination of operating at higher frequencies (e.g., mm-waves, THz) to achieve wider bandwidth for providing very high data-rates for the advanced applications [9, 10]. However, operating at higher frequencies results in multiple undesired effects such as, blocking, atmospheric absorption, propagation losses, creating a bottleneck due to scarce resources (spectrum, energy, etc.). Further, following the present trends, it can be envisioned that within this decade, scenarios will be prevalent in which there will occur a seamless blending of both, the virtual and the real world, and hence, it is clear that a blind move to operate at higher frequencies will only create multiple performance issues, and at some point, there will be no solution(s) to the problems.

An alternate solution appears in the article by Shannon [8] which, in addition to the Level 1 (technical) problem, also identifies the level 2 (semantic) problem, and level 3 (effectiveness) problem. Specifically, as opposed to the level 1 problem, which focuses only on the transmission and correct reception of the data (bits/symbols), the level 2 problem aims at the semantic (meaning or context) exchange of the transmitted data, while the level 3 problem focuses on the effective exchange of the semantic information. The earlier wireless network generations operated effectively under the umbrella of the level 1 problem; however, with the 6G technology's vision of providing a network to enable the pervasive intelligent services requiring emphasis on the effectiveness and the sustainability, inclusion of semantics becomes mandatory. Thus, rather than focusing on only 'how' to transmit, the aim in the present and future scenarios must be 'what' to transmit [11]. With this, communication will occur with an aim to convey the exact meaning or for accomplishing a goal simultaneously considering the impact of received data on the meaning/context interpretation. which was initially the transmitter's intention or to accomplish a common goal. Further, this will aid in identifying only that information which is absolutely required for recovering the meaning which was the transmitter's initial intention or for accomplishing one specific goal. Also, a combination of interpreting knowledge and reasoning tools with AI/ML techniques will pave way for building the semantic learning strategies to enable the existing AI/ML methods for achieving much better interpretation capability(s). The 6G technology in conjunction with these semantic methods will be able to use efficient learning techniques enabled via semantics at the network edge simultaneously enabling 6G wireless networks in improving the performance. Hence, incorporating semantic communication within the intelligent wireless networks will present a major deviation from the communication which is currently focused only to guarantee that the data which is received is correct, disregarding the context/meaning which is conveyed by the data. This will further open doors to new services of semantic nature which will be able to support all such applications involving knowledge sharing between all the interacting parties. Further, in addition to provisioning only interactions between human-to-human, these next-generation applications will also serve interactions of the type human-to-machine and machine-to-machine. Specifically, the aim of such semantic services will be to ensure a connection which seamlessly intertwines the multiple natural and AI techniques thereby, offering intelligence as a service.

In this article we put-forth the idea of incorporating semantic communication within the intelligent wireless networks. The motivation here stems from the fact that, if learning is inductive then, only data correlation is relevant whereas, meaning of the data loses importance. However, the human brain uses observations/meanings, and learns from previous actions to make effective and complex decisions within a very small span of time simultaneously ensuring energy consumption which is sustainable. A similar technique must be adopted in the next-generation wireless networks by incorporating the level 2 and level 3 aspects as the key components of the network design. This will ensure that only the relevant data, which conveys the information that is intended by the transmitter, can be filtered out thereby, ensuring that a pre-defined goal can be identified and met. Further, by not considering the irrelevant data will help in significantly reducing the bandwidth demands, latency and energy. Therefore, goal-oriented and semantic communications will be the key to explore meaning of the received data wherein, the success of task execution at the receiver (level 3 problem) will assume importance as opposed to merely achieving an error-free communication (level 1 problem).

This survey is one of the first to discuss state-of-the-art advances with a focus on semantic communication and related issues for intelligent wireless networks. Further, towards the end of the survey, several important current and future research challenges are presented followed by the proposal of the corresponding research directions. The main contributions of this study are as follows.

- Discussion and exploration of the state of art advancements towards obtaining a semantic communication architecture relevant for the wireless networks.
- Devising a taxonomy, as shown in Fig. 1, of the semantic communication enabled wireless networks via crucial enablers, use-cases, AI/ML schemes which are emerging, and technologies pertaining to communication, networking and computing.
- Presenting and discussing multiple current and future research challenges, and their possible solutions with an outlook towards enhancing research in regard to semantic communications for intelligent wireless networks.

The rest of the article is organized as follows. Section II details the need for incorporating the semantic and effectiveness problems within a novel architecture for facilitating efficient designs for cross-layer interactions with simultaneous focus on goal-oriented communications. Here, we detail the semantic architecture, and source and channel coding models. In Section III, we detail the cross-layer interaction for semantic communications enabled systems. Section IV describes the various state-of-art applications of semantic communications. Section V presents the various research activities which have been conducted concerning the semantic communications in wireless networks. The open research challenges with potential directions are presented in Section VI. Finally, the conclusions are presented in Section VII.

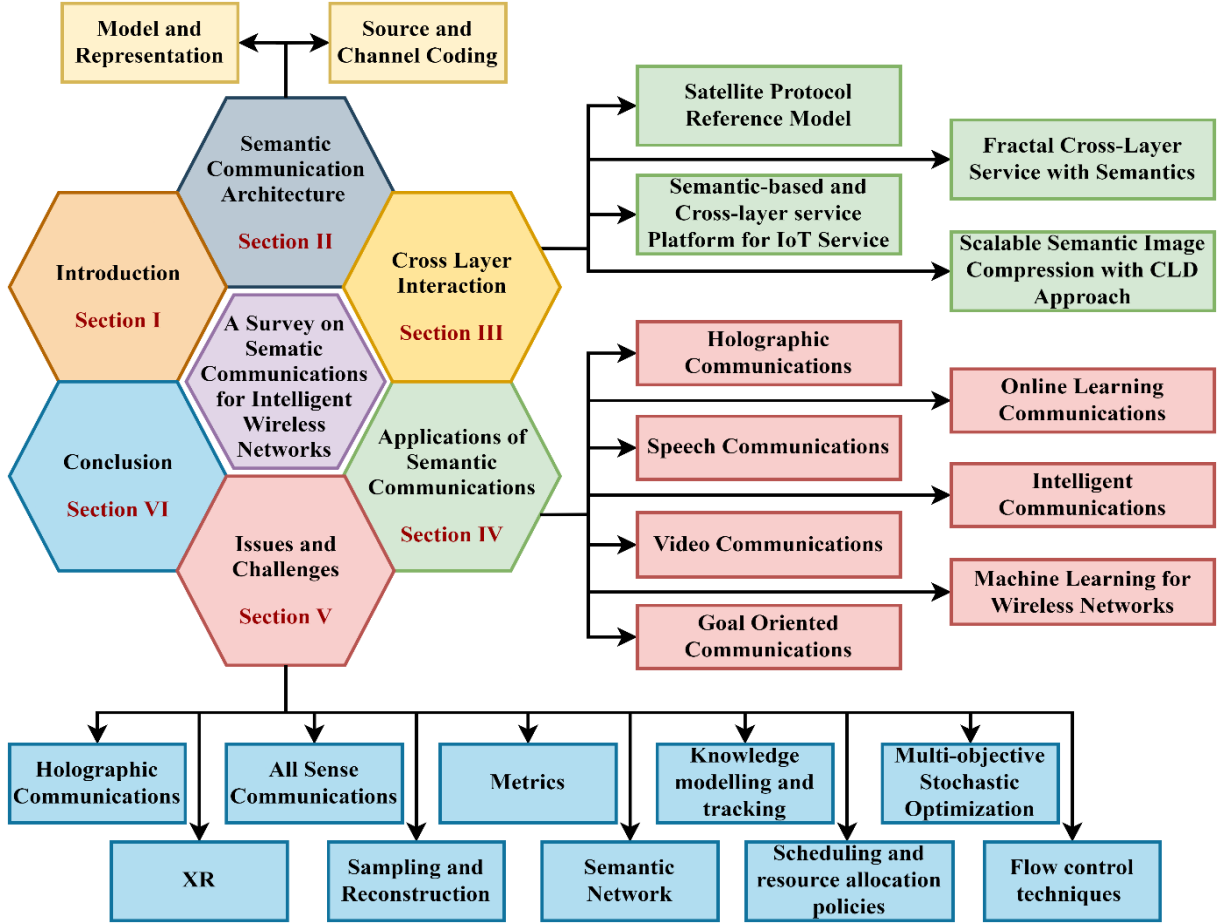

Fig. 1. Taxonomy of the Semantic Communications for Intelligent Wireless Networks survey.

## II. Semantic Communication Architecture

With a move towards the 5G technology, wireless networks are evolving towards a system which involves both, communication and computation wherein, edge cloud supports the tasks requiring communication, computation, and control. In effect, these advanced tasks are equipped to sense, compute, control, and actuate, so as to lay a foundation for the intelligent machines. The 6G wireless networks will further ensure a drastic change by significantly advancing the infrastructure for communication and computation. This will mainly be a result of the increasing pervasive use of the AI/ML tools within all layers of the network which will need a joint coordination of the resources required for computation, communication, and control. As a result, there will be a tremendously fast accumulation of the data which will require almost ideal filtering, transmission, and processing via intelligence of both, natural of intelligent type.

Hence, considering the limits on the resources which are available, the key challenge will be the designing of a new network which is able to deliver the next-generation intelligent services without requiring an exponential increase in the capacity, computation, storage, and energy. Also, observing the capacity limits already reached due to the existing and emerging service requirements, it is evident that operating at higher data-rates to achieve larger bandwidths is not the solution to cater to the needs of the advanced intelligent services. As an alternative, to ensure that the wireless network is qualitatively more efficient, instead of efforts towards increasing the resources, there is a need to ensure that the network is capable of more intelligently utilizing the limited resources. The solution to the aforementioned challenge exists in the work conducted by Shannon [8] which, in addition to the Level 1 or technical problem aiming to only accurately transmit the data, articulates the Level 2 or semantic problem focusing on the precise meaning conveyed by the transmitted data, and the Level 3 or effectiveness problem which overviews the impact of the received meaning in a desired manner. The incorporation of the aforementioned three levels together will ensure that the new services demanding an interconnection of the humans and the machines will possess various intelligence degrees.

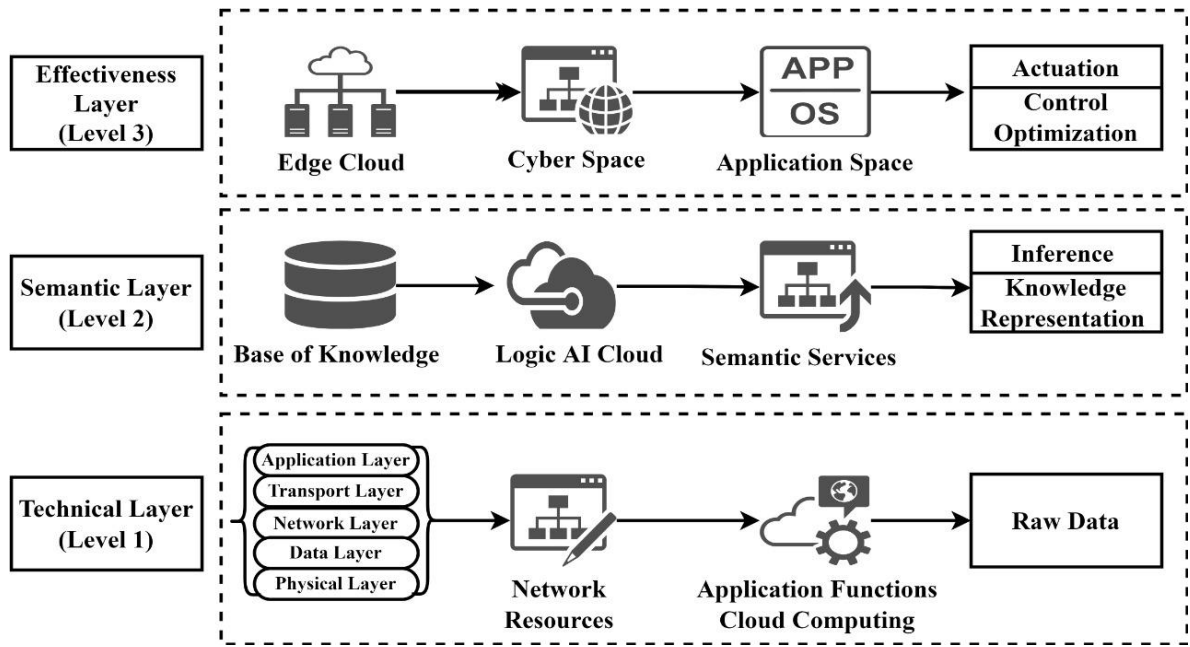

Fig. 2. The new three levels architecture for intelligent wireless networks.

In addition to the technical level, which exists at the bottom of current communication protocol stack, new stack will also include the semantic and the effectiveness levels. With the aforementioned, the new architecture, shown in Fig. 2, consists of a protocol stack which includes all the three levels with Level 1 occupying the bottom layer. In 6G technology, in addition to NFV and SDN, the Level 1 layer will rely heavily on network virtualization which will be able to distribute the tasks involving communication and computation to the virtual machines [12], and this can be further coordinated efficiently through Level 2 and Level 3. On top of the Level 1 layer, the Level 2 layer is placed which will provision such services which require and have a set role for semantics. The Level 3 layer placed on top of Level 2 will ensure that the lower levels function in a coordinated manner so that the available resources can be optimized in their usage and the important performance parameters also achieve the desired values. For instances which do not require any Level 2 aspects, Level 3 will be able to have a direct interaction with Level 1. Such a new protocol stack will enable the layers to conduct an effective cross layer interaction which will be key to allocate the available resources. Specifically, Level 2, if implemented accurately, will ensure that the data which has be transmitted will be interpreted correctly by the receiver irrespective of whether the entire data is decoded without errors. This will in turn offer significant improvement in the performance levels as the communication will now depend on the sharing of a-priori information between the transmitter and the receiver. In regard to Level 3, the aim will be to ensure that communication occurs to achieve a specific goal via correct implementation of the optimized resources with least latency. For this, Level 3 will use the resources available at Level 1 and will coordinate with Level 2, if and when required.

In view of the Level 1 problem, multiple studies have presented alternate methods from an information theory viewpoint which are consolidated in [13, 14]. There also exist studies in regard to semantic communication system [11, 15-30]. In what follows, we present a summary of the most recent studies, and in Table 1, we present the consolidated review of all the studies which exist in literature in regard to the semantic communications.

In view of extended connectivity, authors in [11] have extended the scope of semantic communications to design both, the objectives and the constraints. Authors have encompassed semantic information of transmitted data for any application/use case, and have provided a platform to obtain semantic-aware connectivity solutions via a semantic-effectiveness plane. This enhances the existing stack of protocol by provisioning interfaces which follow standards so as to enable the filtering of data and to directly control the function(s) at all the layers of protocol stack. The authors have demonstrated that the semantic-effectiveness plane replaces current architecture in wireless networks with a framework which demonstrates performance enhancements. Lastly, authors have presented open research problems in regard to cross-layer recovery mechanisms, and security. In [15], the authors have proposed and analysed an end to end framework for communication via semantics which incorporates issues of inference of semantics and communication via the physical layer. The authors have considered the semantic aspect by considering the similarities which exist between individual words. The authors in [16] have envisioned the data semantics as foundation for the process of communication. The authors have pointed out that such a foundation will ensure that data is generated and transmitted with an aim of goal-oriented communication. By taking advantages of the semantics empowered sampling and communication policies, the authors have shown that significant minimization of both, reconstruction error and actuation error cost can be achieved in addition to the generated number of uninformative samples. In [17], the authors have used a deep neural network (DNN) for learning and jointly encoding a semantic channel encoder under the consideration of similarities which exists between the complete sentences. As a key aspect, the study focuses on recovering the transmitted message's context/meaning instead of trying to prevent data (bits/symbols) errors. The authors have proposed a novel metric namely, sentence similarity for justifying performance of semantic communications. It is shown that, compared to exiting communication technique, the proposed method achieves higher performance.

The authors in [18] have designed a deep learning (DL) based system for semantic communication in regard to speech signals. For improving speech signals' accuracy to recover, the proposed system operates on an attention mechanism by resorting to the use of a squeeze and excitation network. For enabling system to tackle dynamic channel environments, the authors have proposed a generic model for coping with multiple channel conditions without requiring any re-training. It is demonstrated that proposed system offers better performance compared to traditional communications, and also shows higher robustness to variations in the channel. In [19], the authors have proposed a model to investigate the issue of audio semantic communications over the wireless networks. The model uses semantic communication methods for transmitting, to server, audio data which is large sized through wireless edge devices. For extracting semantics from audio data, authors propose an auto-encoder based on wave to vector architecture which uses convolutional neural networks (CNNs). For further improving accuracy of extracted semantic information, the authors have implemented federated learning (FL) technique considering many devices and one server. It is demonstrated that new technique converges efficiently and is able to minimize mean squared error of the audio transmission by approximately 100 times in comparison to the traditional schemes of coding. The authors in [20] have proposed a route to boost network capability(s) method in view of enabling 6G wireless network. The authors have (i) conceived a novel semantic framework, named as semantic base, and (i) established an intelligent and efficient semantic architecture which integrates AI and network technologies for enabling intelligent interactions among various 6G communication objects. The authors have also presented a survey of recent advances in semantic communications simultaneously highlighting potential use cases.

In [21], the authors have discussed the relationship between semantic communications and IoE, and have introduced the basic models and fundamental components of semantic communications. The authors have discussed the major limitations of the point-to-point semantic communications, and have hence put forth that a knowledge sharing and resource convergence enabled semantic communication networking is ideal to support the future massive scales of IoE systems. In addition, the authors also discuss the basic components of the semantic communication networking system, and a federated edge intelligence enabled semantic communication networking architecture is investigated as a case study. The results demonstrate the potential of semantic communication networking to further minimize the resource demand(s) and result in an improvement of the semantic communication efficiency. Lastly, the authors have discussed the open problems for future research. The authors in [22] have explored the opportunities which are offered by the semantic communications for the next generation wireless networks. Specifically, the authors have focused on the benefits of semantic compression, and have presented a detailed new architecture which aids in enabling the semantic symbols in view of effective semantic communications. The authors have also discussed the theoretical aspects and have designed the objective functions which aid in learning the effective semantic encoders and decoders. Lastly, the authors demonstrate promising results considering transmission of text scenario when transmitter and receiver converse in varied languages. In [23], the authors have reviewed the classic semantic communication frameworks following which they have summarized the key challenges which hinder the popularity of semantic communications. The authors observe that few semantic communication processes result in excessive resource consumption and are hence inefficient. As a solution, the authors have proposed a new architecture which is based on intelligence via federated edge for supporting a semantic enabled network which is resource efficient. The proposed architecture provides security to the model related information by coordinating through the intermediate results. The obtained results demonstrate that proposed architecture is able to minimize resource consumption simultaneously demonstrating significant improvement in the communication efficiency. The authors in [24] have introduced the concept of physical elements including genie, which also features with the 6G technology concept. The authors have focused on genie realization in view of intelligent transmission/access within the 6G networks in conjunction with semantic information theory and AI joint transceiver design. Further, AI is integrated with transmitter and receiver design including multiple granularities, and a complete end to end AI transceiver is designed to optimize parameter via DL within estimation of channel, detection of signal, etc. The authors demonstrate that a genie enabled wireless system operates with high intelligence and demonstrates better performance compared to manual control. Lastly, the authors present a related comprehensive survey and provide a scope for future research with relevant suggestions.

In [29], the authors have incorporated the data semantics in a system of networks. Further, the authors have developed advanced semantic metrics, an optimal sampling theory, semantic compressed sensing techniques, and generation of data which is semantic aware, coding for channel, etc. It is demonstrated that proposed architecture enables to generate appropriate data amount and to transmit correct content to apt place at the correct duration. This is possible via a conjunction of process involving redesign of data generation, transmission and utilization. Overall, authors have concluded that semantic networks require a combination of concepts and tools which are developed separately. Lastly, as a scope of further research, authors have identified the following issues: scheduling, random access, packetization, and operation at higher levels. The authors in [30], have implemented semantics for solving the issues of spectrum and energy by proposing a framework for transmission under high semantic fidelity. The authors have introduced the framework for semantic fidelity for improving the efficiency following which they have introduced transforming of semantics to convert input to semantic bits/symbols. In comparison to the existing transforms, the proposed transform incurs data loss which in turn saves on bandwidth simultaneously provisioning high semantic fidelity. The authors have conducted performance evaluation under the consideration of semantic noise and have presented an audio transmission case study for the evaluation of effectiveness. Lastly, the authors have discussed various applications and multiple open research problems.

In the following sub-sections, firstly, we present a detailed description on the semantic communications model and representation following which, the source- and channel-coding are described. The reader must note that in what follows, the term 'semantic data' refers to the meaning associated with the data, and the term 'syntactic data' refers to the data's probabilistic model which is used to encode the data [24].

Table 1. Summary of various existing studies on semantic communications.

| Reference | Aim | Problem addressed | Method | Major Results | Future Scope |
|---|---|---|---|---|---|
| [11] | Extend scope in regard to designing of the of the aim and constraints. Also, include the transmitted data's semantics for any application or use-case. | Provision a framework to obtain solutions which are enabled by semantics via a semantic-effectiveness plane. | Augment the existing stack of protocol through interfaces which follow standards. | Semantic-effectiveness plane replaces the current version by a framework which regularly improves and extends both, the system and the standards. | Cross-Layer Recovery Mechanisms, Security |
| [15] | Propose a framework for communication which accounts for context of the transmitted data under the consideration of a channel which is affected by noise.. | Characterize optimal transmission policy(s) for reducing end to end mean semantic error | Formulation of the communication issue in the form of a Bayesian game. Investigation of conditions for which there is existence of the Bayesian Nash equilibrium. Consideration of an online communication case wherein, actions are taken sequentially so as to form beliefs about other party. Finally, evaluate sequential game, structure for system of belief and profiles of strategy. | Semantics of word can be used to assess performance of communication. Also, the beliefs of transmitter and receiver influence the optimal strategies which are adopted for communication. | Investigate the trade-off between encoding and decoding optimality functions and words amount. Consider the cases of incorrect data and/or manipulated received data. |
| [16] | Capitalize on semantics driven sampling and communication policies. | Apply a method which is new in structure and has synergy to ensure that receiver must perform real-time | Semantic-enabled system with foundations of goal-oriented data aggregation, and | Major minimization in errors owing to reconstruction and actuation error. | Multi-objective Stochastic optimization, Semantics-aware Multiple Access, Goal-oriented |
| | | reconstruction at the source to enable remote actuation. | transmission and utilization of data, The proposed model implements the attributes of data's context. To tailor method such that it meets the needs of networked applications in view of achieving specific aims. | | Resource Orchestration, Semantic metrics |
| [17] | Proposal of DL enabled semantic system for transmission of text. | Model increases the capacity of the system, and reduces errors of semantic nature by recovery of meanings of the sentence(s) meaning. | Transfer learning ensures that proposed model is applicable to various communication environments, and accelerates the training process of the model. A new metric, namely sentence similarity, is initialized | Model shows robustness to variations in the channel and achieves better performance. | - |
| [18] | Recovery of transmitted speech signals in in a semantic system, which reduces error at semantic levell | Designing a DL enabled semantic system for speech signals. | Improvement of speech signals' recovery accuracy via a model based on attention mechanism which uses squeeze and excitation network. Facilitate model to operate under dynamic channel environments, and multiple channel conditions without retraining. | Model is robust to channel variations. | - |
| [19] | Investigate audio semantic communication issue in wireless networks. | Development of real time audio semantic system via which large audio data can be transmitted to a server by the wireless devices. | The model for audio communication is trained using FL. | The proposed algorithm effectively converges, and reduces mean squared error (MSE) of audio transmission by approximately 100 times. | - |
| [20] | To ascertain evolution towards wisdom evolutionary and primitive concise network. | Propose route to boost network capability(s) method. | Conceive a new semantic representation framework named as semantic base. Establish a semantic communication network | Application cases and future scope are detailed. | Semantic enabled intent driven networks, fundamental limits of semantic communication, development of base of knowledge representation |
| | | | architecture in view of efficiency and intelligence. | | |
| [21] | The relationship between semantic communication and Internet of everything is discussed. Basic models and fundamental components of semantic communication are introduced | It is argued that knowledge sharing and resource convergence based semantic communication networking would be ideal for supporting the future massive scales of Internet of everything systems. | Federated edge intelligence-based semantic communication networking architecture is considered as a case study. | Semantic communication networking has the potential to further reduce resource demand and improve efficiency of semantic communication | Metrics for semantics, Sampling and reconstruction, Resource allocation methods |
| [22] | Explore possibilities offered by semantic communications for next generation wireless network including semantic compression | New architecture to enable semantic symbols representation learning for ensuring efficient semantic communications. | Discuss theoretical aspects following which objective functions are designed to aid the semantic encoders and decoders in learning. | For varied language speaking source and receiver, results are demonstrated to be promising. | Extend the gain by application to multiple modal and data hungry applications. |
| [23] | Survey frameworks for semantic communication and detail the major issues which make it unpopular. | Propose new architecture enabled via federated edge intelligence for supporting semantic enabled network with resource efficiency. | Proposed architecture permits to offload large computations to edge servers. | Few semantic communication processes are demonstrated consume more resources and result in inefficiency. | Knowledge Evolution Tracking, Network-level quality of experience Quantification, Capacity of Semantic-aware Network |
| [24] | Introduce the concept of four physical elements viz., man, machine, object, and genie, which also features with the 6G technology concept. | Focus on genie realization in view of intelligent transmission/access within the 6G networks in conjunction with semantic information theory and AI joint transceiver design. | AI is integrated with transmitter and receiver design including multiple granularities. A complete end to end AI transceiver is designed to optimize parameter by DL method. | A genie enabled wireless communication link operates with higher intelligence and performs better compared to manual control. | The genie concept can be used at the following layers: MAC, RRM, network, and application. |
| [25] | Study the manner in which semantic data can be measured quantitatively. | Investigate a model's performance in view of semantic data compression and reliable communication. | Relate the proposed method to data measurement which is statistical in nature. | There exist counterparts to Shannon's source and channel coding theorems. | Semantic coding algorithms with efficiency. |
| [26] | Extend classical source coding theorem to include semantics at transmitter. | Manner in which semantic relationships existing within an information rich source can be exploited. Realistic semantic compression algorithm to exploit graph structure of a | Define capacity of a semantic transmitter. Show that the semantic transmitter capacity is equal to the data mean semantic entropy. | By using semantic relations between transmitter symbols, higher rate of lossless compression may be achieved. | - |

| | | shared base of knowledge. | | | |
|---|---|---|---|---|---|
| [27] | Initiate a study on communication aims | Present two definitions viz., meta-goal, and syntactic goal. | Formalism is able to capture multiple commonplace examples of communication goals. | For few technical conditions, meta-goals with syntactic versions, are universally achievable. | Identify the specific resource whose usage the language wants to optimize. |
| [28] | General theory of goal oriented communication | Proposed model accounts for all the aims. | Identify concept of sensing which will allow goals to be achieved even where there is any misunderstanding | Any communication aim is modelled mathematically via a referee. When user senses progress, communication aim can be achieved despite initial misunderstanding. Sensing may be required to overcome initial misunderstanding. | Focusing on the aim helps in detection and possible correction of the misunderstanding. |
| [29] | Incorporate the data semantics to communicate and control in a networks. | Develop advanced semantic metrics, an optimal sampling theory, and semantic compressed sensing techniques. | Architecture to enable generation of correct data amount and transmit apt content to receiver. | Combine concepts and tools which involve multiple domains. | Scheduling, Random access, Packetization, Operation at higher levels |
| [30] | Implement semantics for solving the issues of spectrum and energy. | Propose a framework for transmission under high semantic fidelity. Introduce the framework for semantic fidelity to improve the efficiency | Introduce transformation of semantics to convert input to semantic bits/symbols. Proposed transform incurs data loss to save on bandwidth simultaneously provisioning high semantic fidelity. | Performance evaluation under the consideration of semantic noise. Present an audio transmission case study for the evaluation of effectiveness. | Discuss various applications and multiple open research problems. |

### i. Model and Representation

A semantic communications model is efficient if it ensures the correct communication between source and receiver provided that the receiver (i) recovers true meaning of the data sent by the transmitter from the data which has been received, and (ii) increases the related knowledge from the data which has been received. The aforementioned implies that there will occur a semantic equivalence when the receiver infers the same meaning from the received data as was intended by the transmitter. Such an inference requires a model which deviates from the level 1 aspect of the Shannon's theory in the following manners: (i) semantic content decides the data amount rather than the probability of symbol generation for encoding the data, (ii) exact content of the data is important rather than the mean data which has an association with the most likely information that can be transmitted, and (iii) in addition to the information which is conveyed by the data depending on the data itself, it also depends on the knowledge level which is available at the transmitter and the receiver.

Further, as semantic communication is related to the meaning of the data, a system of knowledge will be required which will formally represent knowledge associated with the semantics of the data. This representation of the knowledge, which is the base of AI mechanisms [31], will aim to efficiently represent and interpret the data via appropriate definitions of the information carried by the data thereby, also possibly creating new knowledge. Therefore, a base for knowledge representation and interpretation will be required at the transmitter and the receiver which may differ from one another. It must be noted that for the base of knowledge, incomplete description will be a key feature which will distinguish it from a database. The incompleteness will arise mostly owing to the constraints posed by computations since to completely finish reasoning will consume large amounts of time. Also, data will be correctly received only when the associated meaning is same as was intended by the transmitter, or it results in a value addition of the base of knowledge. Further, the received data will be depended only on the semantics rather than the syntactics which implies that the data will have multiple encoding options which could give rise to similar semantic meaning which is an open problem for research.

### ii. Source and Channel Coding

The multiple level communication flow diagram is shown in Fig. 3 which shows the three layers in regard to three communication levels viz., level 1, level 2, and level 3. At layer 3, there will be a transmitter and a receiver which will interact with one another via an environment. The nodes may correspond to humans or machines, also referred to as the 'agents', which will operate as detailed in [31]. These agents may also be rational by implementing an interaction which may include data exchange, control, sending, etc. Specifically, for interaction, the transmitter will generate a message $m \in M_T$ (where $M_T$ represents a transmitter alphabet), following the rules set by the base of knowledge which exists at the transmitter, such that it will deliver the intended meaning to the receiver. Further, for physically conveying the data to the receiver through the channel, the data ($m$) will be converted to a symbols sequence, $s \in S$ (where $S$ represents the alphabet of the symbols), which will then be converted to physical signal which is apt to propagate over channel. Also, it must be noted that the mapping from $M_T \in S$, which is denoted as $s = f(m)$, will always not be a one-to-one mapping, and could also be a one-to-many mapping since any data is represented via many symbols which convey similar meaning. The aforementioned will result in ambiguity which in one of the key research challenges in the domain of Natural Language Processing (NLP) [32].

Following the Shannon's theorem, a source encoder is required to translate $m$ to $s$ such that redundancy within the data can be minimized. This will be followed by a channel encoder, which will implement only the most required redundancy, which will aim to increase the reliability of the communication. It must be noted that this combination of the aforementioned encoder for source and channel will form syntactic encoder since it will only focus on correctness of data and not the related meaning (semantics). Lastly, complete sequence $s$ is converted to physical signal which is compatible with the channel being used for propagation.

Consider a semantic encoder in which the transmitter is random and transmits the data with a probability $pM_T(m), m \in M_T$. Therefore, following the Shannon's theory, the data entropy will be given as

$$H_T(m) = -\sum_{m \in M_T} pM_T(m) \log_2 pM_T(m) . \tag{1}$$

Further, following the study in [25], the probability that a transmitter is able to transmit a symbol $s$ can be evaluated as

$$p_T(s) = \sum_{\substack{m:s=f(m) \\ m \in M_T}} pm_T(m) . \tag{2}$$

where, the probability of a transmitter being able to transmit a message $m$ is denoted by $pm_T(m)$. Thus, for any symbol $s_i$, the corresponding semantic information is given as

$$H_T(s_i) = -\log_2 p_T(s_i) . \tag{3}$$

and for the symbols transmitted by the transmitter, the corresponding semantic entropy is given as

$$H_T(S) = -\sum_{s_i \in S} p(s_i) \log_2 p_T(s_i) . \tag{4}$$

Following the Shannon's theorem, for only level 1 problem, two differing entropy, $H_T(M)$ and $H_T(S)$, can be defined as

$$H_T(S) = H_T(M) + H_T(S/M) - H_T(M/S) . \tag{5}$$

where, $H_T(S/M)$ is entropy of $S$ conditioned by $M$ whereas, $H_T(M/S)$ is entropy of $M$ conditioned by $S$. However, in semantic communications, the aim of source coding is to preserve the meaning (semantics) of the data rather than the bits/symbols sequence which is generated by the transmitter. With this understanding, $H_T(S/M)$ denotes the semantic redundancy since it will differ from zero only when many bits/symbols which are associated to a similar message exist, and $H_T(M/S)$ denotes the ambiguity in semantics since it will differ from zero only when many contexts associated to similar bits/symbols exist. Further, the study in [26] has shown that for such a scenario, there will exist a semantic encoder which will require, on an average, $I(M;S)$ number of bits for encoding data which is transmitted by the transmitter. Hence, the mutual data which exists between the transmitted messages and the transmitted symbols is denoted as

$$I(M;S) = H_T(M) - H_T(M/S) = H_T(S) - H_T(S/M). \tag{6}$$

Also, few practical semantic transmitter encoders proposed in [26] exploit the shared knowledge between the transmitter and the receiver.

Overall, as the main aim, semantic encoder detects and extracts meaning from the transmitted data and then compresses or removes information which is not relevant. To do so, initially, through the knowledge existing at transmitter and receiver, the encoder will have to identify the related entities from the transmitted image/text following which, it will have to infer the closest relationship in accordance with a common model.

At the receiver end, the signal $r$ which is received is decoded syntactically for generating $s'$ symbols sequence. Next, depending on the base of knowledge which exists at the receiver, $s'$ is interpreted to generate the message $m'$. From the aforementioned, having received the signal $r$, aim of semantic decoder is the recovery of message $m'$ which is same as transmitted message $m$. Equivalence here will imply that $m'$ and $m$ deliver exact same meaning and not necessarily same structure. Hence, decoder will have to interpret the information which is transmitted by the transmitter, and then recover the signal which is received in a form which is understood by user at receiver. Further, decoder will aim to evaluate satisfaction level of user at the receiver based on which it will decide the success of the semantic data which has been received.

However, as in the syntactic communication case, there will be sources of errors in the semantic communications case also. The error will occur at the semantic level if $m'$ is unequal to $m$, and at the syntactic level if $s'$ differs from $s$. Further, at the syntactic level, the main source of error will random noise or interference during the transmission or propagation of the signal whereas, at the semantic level, the error will mainly occur if the base of knowledge at the transmitter and receiver differ or if there is misinterpretation of the data. This implies that semantic layer is reliant on the syntactic layer wherein, multiple errors in received signal decoding may affect data recovery. However, this does not necessarily imply that there will errors in the interpretation of the data i.e., errors at the syntactic level do not necessarily result in errors within the semantic level since, even with few errors, the interpreter at the semantic level can still recover the meaning from the received data by exploiting the base of knowledge which exists at the receiver.

Also, there could occur no errors at the syntactic layer but at the semantic layer when the received message is decoded correctly; however, it is not interpreted appropriately due to the difference in the base of knowledge which exists at the transmitter and the receiver. The aforementioned is demonstrated as follows: let there be a channel which is modelled such that the conditional probability is $p(r/s)$ to receive $r$ when $s$ has been transmitted. For semantic communication, the aim is to recover the meaning and not the symbol $s$ which was transmitted, and hence, a semantic decoder can be used which will select $m'$ such that it can increase (or maximize) the a-posterior probability which is conditioned to the received symbol, and is given as

$$m' = \arg\max_{m:s=f(m)} p(m/r) = \arg\max_{m:s=f(m)} \sum_s p(m,s,r). \tag{7}$$

Further, using property of Markov i.e., $p(r/m,s) = p(r/s)$, decoding equation is re-written as [25]

$$m' = \arg\max_{m:s=f(m)} \sum_s p(r/s) \, p(s/m) \, p(m). \tag{8}$$

In the aforementioned, the values of $p(m)$ and $p(r/s)$ are known, and optimizing the complete performance of the system will involve a searching of that $p(s/m)$ which will reduce the probability of the semantic error under the multiple physical layer constraints. In fact, the $p(s/m)$ value has a key role within the functioning of the semantic encoder wherein, in the case of few errors at the syntactic level will imply that the performance can be significantly improved via semantic decoding as most of the received data can be corrected using the base of knowledge at the receiver using natural of artificial intelligence.

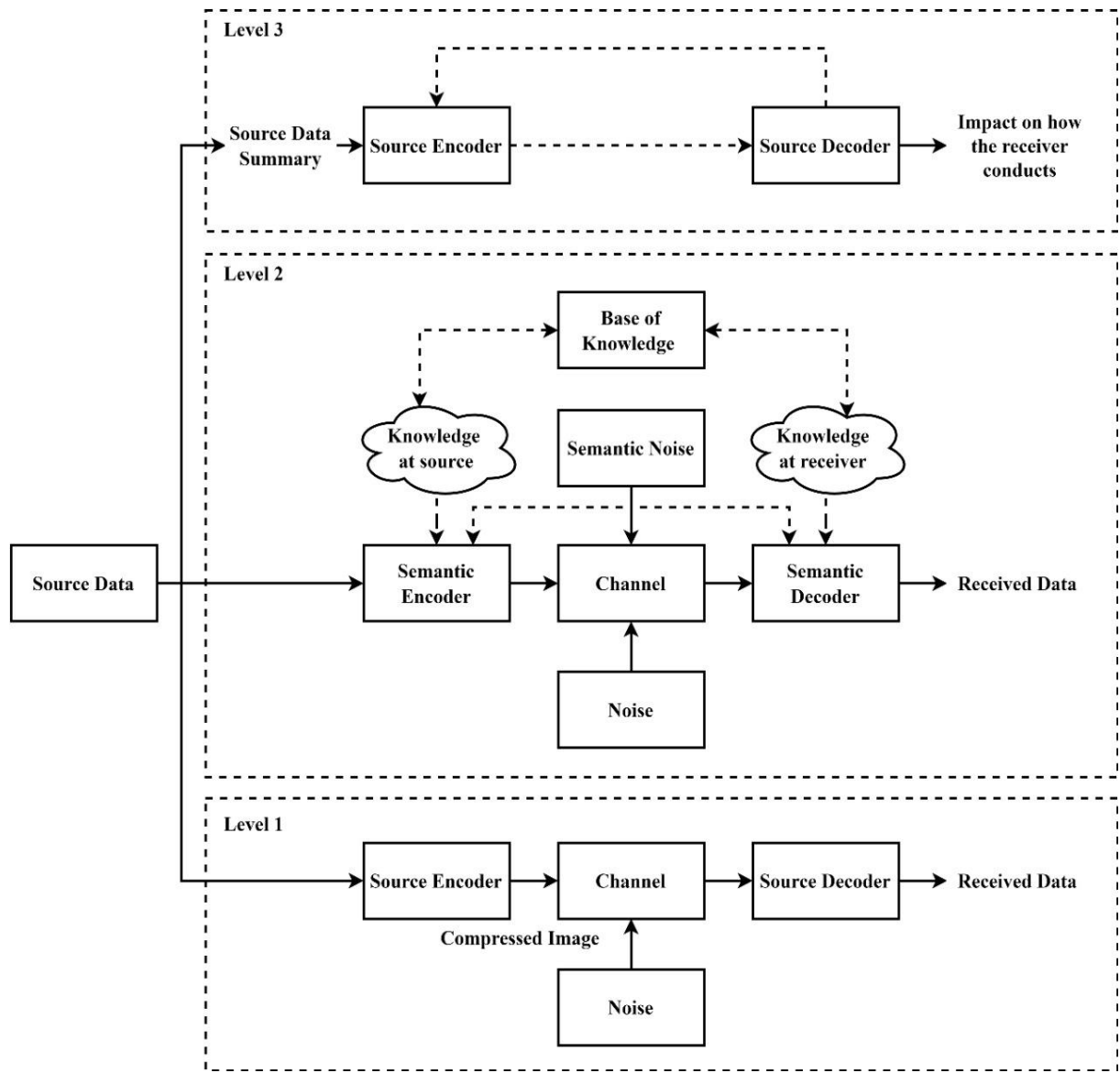

Fig. 3. Multiple level communication system involving the three levels of communications.

## III. Cross Layer Interaction

With increased usage of real-time applications over the wireless networks, high performance and efficient quality of service (QoS) are essential to the end user. Hence, existing TCP/IP method will not suffice in view of meeting the increasing heterogeneous demands, and therefore, much research must be focused on various cross-layer design approaches for increasing the overall performance and QoS by allowing data sharing across the various layers.

To meet the increasing demands, timely exchange of the data across layers, and periodic reconfiguration of modes due to increased mobility cross-layer designs (CLD) have been proposed [33]. It is a common misconception that CLD destroys the traditional 5-layer architecture; however, CLD also works in conjunction with the layered architecture to enable cross-layer communication between non-adjacent layers. Using CLD, the layers may share parameters and internal details to enable effective troubleshooting of the root cause with lesser processing power and reduced cost. Also, CLD can achieve reduced latency, increased throughput, and lesser error rate which are vital for next-generation wireless communication applications [33].

The coordination plane is a model used to showcase the issues which can be solved by CLDs and the corresponding advantages. Here, each coordination plane represents an issue which can be solved by CLD. Four different types of coordination plane are labelled, as shown in Fig. 4.

- **Security:** Layered TCP/IP protocol leads to multiple levels of encryption across various network layers with increased processing and costs.
- **Quality of Service:** The coordination plane emphasizes on increased QoS which is possible by sharing certain information across the non-adjacent layers also.
- **Mobility:** CLD is built to provide seamless and uninterrupted communication in situations of augmented mobility.
- **Wireless Link Adaptation:** This coordination plane emphasizes on reducing the bit-error rate (BER) and channel fading in wireless networks using the CLDs.

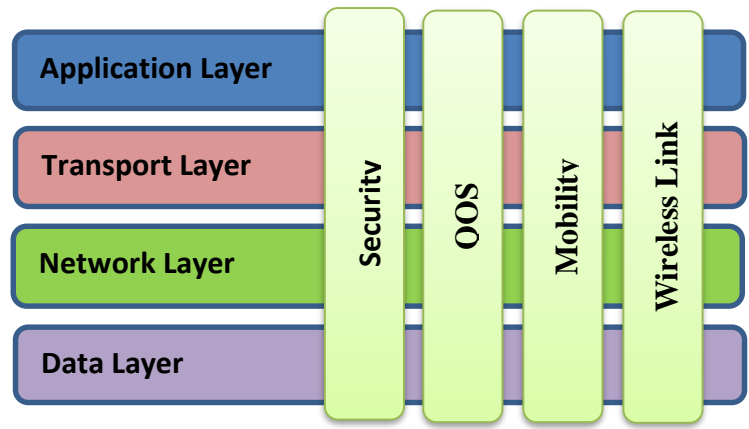

Fig. 4. CLD Coordination planes.

Further, as detailed in [34], a probable solution to CLD is to conduct a cross-layer design to shield the upper layers from operational details of the handover; however, it is also important to inform the upper layers about the handover(s) to enable them to adjust to the handover. In what follows, we present few optimization techniques which could be adopted with CLD enabling the essential communication between different network layers which is required for the optimizations. Specifically, we detail the CLD approaches and the corresponding execution problems that require upcoming investigations.

i. **Satellite Protocol Reference Model**

The European Telecommunication Standards Institute Technical Committee-Satellites Standard Earth Stations and Systems/Broadband Satellite Multimedia (ETSI TC-SES/BSM) has demarcated IP-primarily based satellite system structure containing bottom layer air interfaces [34]. Fig. 5 demonstrations this sort of protocol architecture in which, the bottom layers rely upon the satellite structure for putting into practice and the top layers are standard of the IP suite (satellite-free layers). Such two protocols layers are interrelated due to satellite-free-service access point (SF-SAP) interface, and standardization envisions small range of typical functions which pass SF-SAP. Precisely, those features are deal with resolution, aid management, and visitors training QoS. The mission exists in the implementation of a move-layer technique (i.e., linking satellite system for pc-based and satellite independent layers) which debts for the aforementioned protocol shape and the SF-SAP interface. More specifically, appropriate primitives should be included within the standardization for supporting such a prolonged signalling.

ii. **Fractal Cross-Layer Service with Semantics**

Pervasive assembly, developed platform for service, and precise scientific model of service depending on logic of service in order to yield quick and clever actions are the most significant characteristics of services in the next-generation IoT [35]. IoT can achieve ubiquitous connection, and hence, the manner in which this technology can be used to build cross-layer platform to provide service with active integration and interoperability is a major challenge. For example, in sensing enabled technology, IoT serves as a service oriented system having a core value of "smart services" [36, 37]. In addition, to including common services in traditional internet, IoT services comprise pervasive services in different network setting such as, mobile networks and wireless sensor networks (WSNs), etc. [38]. Further, IoT services are considerable and diversified in nature, and typically face morphological changes, environmental change, outward expansion, business restructuring, altered sharing and interoperability levels problems, and other situational dynamic adaptabilities. These issues present challenges to the IoT service platform, with possible solutions such as, web services, semantics, etc., which aid in achieving services with efficient integration, sharing, discovery, and interoperability [39]. With this, an effective service discovery and personalized delivery can be achieved [40, 41].

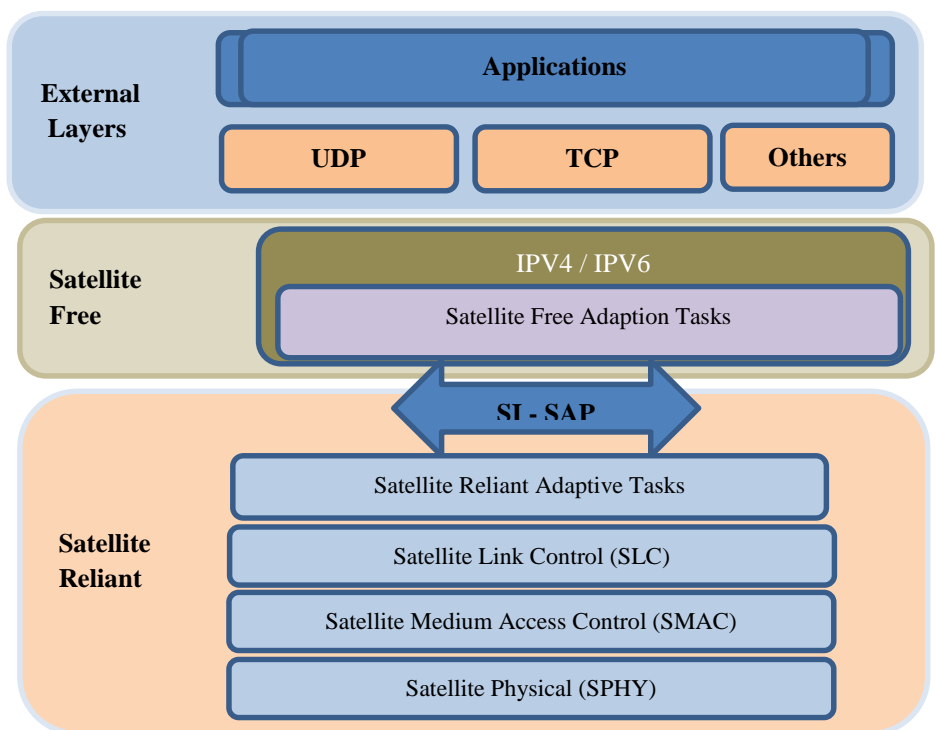

Fig. 5. Satellite protocol architecture by ETSI TC-SES/BSM.

iii. **Semantic-based and Cross-layer service Platform for IoT Service**

In this sub-section, following the comprehensive analysis of web service architecture and semantic based sensor network architecture [42] projected in EU FP7 [43], the new semantic-dependent IoT service architecture is presented in Fig. 6. Such an architecture comprises of two types viz., the service based concept involving both, semantic description and IoT ontology, and corresponding service operation. The architecture includes handler ontology which aids the description of IoT service resource. Finding practical requirements includes service and quality ontology to attain semantic depiction of user's practical needs. Context ontology will be used to confirm semantic annotation on characteristics of context that are attained by context aware computing. Semantic-related characteristics of user helps to meet the user's adaptive and individual needs of the improved service. Further, service ontology creates service, then publishes it, and it exits throughout the entire platform of process [44]. Further, this framework contains service discovery, selection and combination. Service composition process is attained by dividing requirements, and service discovery involves the examining and comparing process on the original advertising services set based on service requester's service requirement. Thus, it could return service set which can meet functional requirements. The service resource can also be organized and then managed by decentralization. Finally, the service selection corresponds to a process of selecting the personalized service depending on service requester. In specific, it needs to select the service discovery and choice strategy to adapt in any dissimilar case, with strategy, service ontology, quality ontology, context ontology, etc., which ensure a semantic assisting role. During course of the outmost part of the framework architecture, there exist privacy, security and trust that offers the basic assurance for the execution of the complete process.

iv. **Scalable Semantic Image Compression with CLD Approach**

In an intelligent society, image compression aims to serve human imaginative, prescient and, MV. Traditional picture compression schemes will not ignore visual fine for viewer. The process of complete decoding of images is necessary earlier to their application to the semantic analysis. These elements make traditional schemes semantically inefficient. Hence, it is advised to compress and then transmit the photo indicators and features concurrently in order to efficiently serve the needs of HV and device imaginative and prescient. A unique semantic scalable image compression approach has been proposed. The method gradually compresses the "coarse-grained" semantic capabilities, the first-class-grained semantic functions, and the photograph alerts as shown in Fig. 7. To make use of the pass-layer correlation among functions and alerts for image, a "pass-layer" context model has been recommended to lessen the information redundancy. This approach ensures enhanced layer capabilities as move-layer appears to predict the distribution parameters. This is possible due to use of the entropy form of "decrease-layer" functions. Furthermore, by considering the region of interest (ROI) compression system, in which the equipments using high semantic records are used and the background are compressed one after the other in order to enhance the compression performance. It has been experimental shown at the CUB-200-2011 and FGVC-Aircraft datasets are efficient in assessment of the methods which one by one compress the picture indicators and capabilities.

Currently, intelligent multimedia system applications square measure taking part in a crucial role in daily life like, in good cities and in intelligent police investigation. It is not possible to process and analyse the increasing image/video knowledge simply via human vision. Therefore, there is surge in application of machine vision techniques that are responsible for the speedy development of decilitre, and DNNs. The major role of these devices is in performing visual investigation by machine. However, it is not possible to substitute human understanding and decision-making by the MV algorithms. Thus, in order to facilitate human-machine interaction, compression techniques have to work for both HV and MV needs.

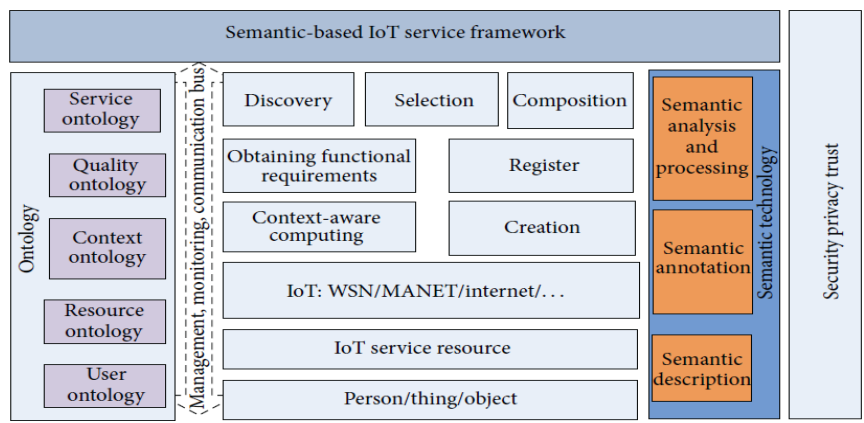

Fig. 6. Framework of IoT service based on semantics.

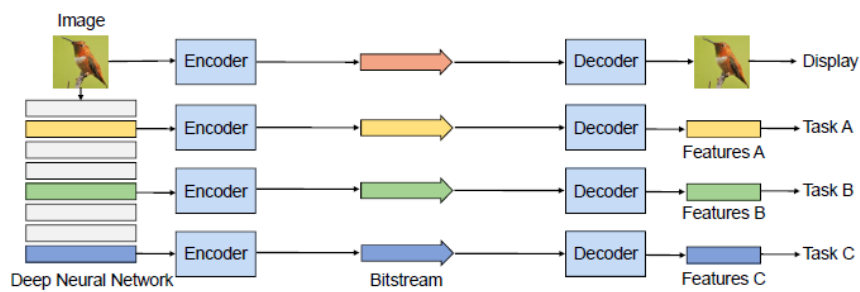

(a)

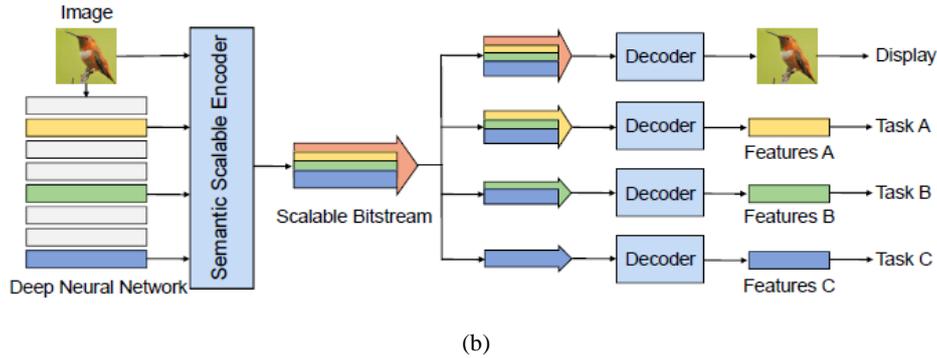

(b)

Fig. 7. (a) Individual compression together. (b) New scalable semantic image compression.

A technique of first compress and then analyse is used in conventional multimedia system. The image signals square measure compressed as bit stream for transmission and storage. In the next step, the receivers on works on the decoded signals to rewrite the bit stream to perform visual study tasks. Further, traditional compression ways primarily target the image quality for HV [45, 46]. The linguistics fidelity isn't thought of in such a scenario. Therefore, compression of artifacts that appears below low bitrates extremely worsen the performance of the linguistics conditions [47]. Additionally, the visual analysis method is sometimes resource overwhelming, particularly using DNNs [48-50]. The result of this is observed in terms of many procedures overhead for "back-end devices or servers". Hence existing compression approaches fail to address matter of "human-machine co-judgment" expeditiously, and so, for enhancing potency of linguistics tasks and cut back, complexness of linguistics analysis, another technique is "analyse-then-compress". In this technique, option square measures computed and compressed are used for analysis in a very simple manner by the receivers. Within the theme of decilitre, options square measure sometimes extracted by DNNs, which might be considered as a heap of "feature extractors". Therefore, options here are even having a "multi-layer structure", and also the learning method of DNN is neglecting task-independent info increasingly [51]. An existence of additional abstract and task-specific info results in higher-layer options that are troublesome to generalize to different analysis tasks.

Recently, intermediate function compression has gained immense interest [52-55] because of the reality that computational load may be transformed to "front-end" whereas, ability to generalize compressed capabilities is preserved. Nevertheless, because of records loss inside function extraction procedure, it is challenging to restructure the unique photographs from the restructured functions, which limits its possibility of application as human imaginative and prescient is also crucial. To help each, human imaginative and prescient, and system imaginative and prescient, and additionally discover a better alternate among the computational load and the generalization ability, the best scenario occurs when, simultaneous transmission and compression of photo indicators and functions occurs. Consequently, receivers may request appropriate features consistent with necessities of analysis obligations and may also request photo indicators for presentation. A powerful method is compression of picture alerts and features for my part, as shown in Fig. 7(a), wherein, generated bit streams are simultaneously transmitted. Nevertheless, these manner desertions the correlation among the functions and the photos, and as end effect, the efficiency of compression techniques may be actually poor.

The preceding sections and this section has made it clear that, in evaluation to a syntactic level conventional receiver which requires re-transmission of the packets which are in errors, on the semantic level, the receiver re-transmits handiest whilst any mistakes occur on the semantic stage. Also, further to the traditional approach of regarding the transmitted bits/symbols sequence corresponding to the syntactic stage, a framework related to the semantic stage might be capable of presenting remarks from the receiver to the transmitter thru the involvement of semantics of the transmitted records. However, the principle mission to be addressed is that of devising such mechanisms in order to be capable of hit upon the errors which arise on the semantic ranges. A possible technique to this problem will be that a semantic orientated feedback stating the requirement of a re-transmission may be sent by means of the interpreter on the receiver to the source generator at the transmitter in case when the meaning of the received statistics is not clear.

Further, in view of pass-layer interactions, there can be a constant statistics trade among the syntactic and the semantic ranges. As an instance, if the received information is being always interpreted and decoded effortlessly within the favoured time body, the interpreter (semantic) on the receiver will ship a comments to the encoder (syntactic) on the transmitter declaring that no longer all the facts desires to be transmitted, hence being able to limit the statistics-rate (bandwidth), and power (strength). Such an interaction between the diverse ranges paves manner to new designs for conversation structures in which, in spite of errors at the syntactic layer, correction can be performed on the semantic layer without the requirement of re-transmission of information. As an instance, inside the 5G communication running at mm-waves wherein blocking off and absorption because of atmospheric consequences is excessive, the transmitted statistics won't reach the intended receiver; however, the usage of appropriately tuned models for prediction, the interpreter at the semantic stage will nonetheless be capable of reconstruct the semantic records.

Overall, the common knowledge which is shared among the transmitter and the receiver enable multiple errors correction without having to retransmit the statistics; however, at increased receiver complexity.

## IV. Applications of Semantic Communications

In this section, we detail the various applications which can be enabled via semantic communications to improve their performance.

### i. Holographic Communications

Holographic communications, a challenging new use-case foreseen for 6G networks, is a technique in which many views of one scene are transmitted to create a hologram at the receiver end. In this regard, semantic communications can play a major role in the advancement of holographic technique by incorporating the semantic aspects which will be shared by the transmitter and the receiver to ensure a common base of knowledge. Also, for serving the next generation use cases such as, wireless brain-to-computer interactions (WBCI), mixed reality (XR), Internet of robots, etc., semantic communications enabled applications will be required. However, to attain the advantages provided by including the semantic aspects, the price to be paid is that of additional computational complexity at the receiver end which will result in being a major bottleneck for few applications. To minimize the delay, semantic communication enabled systems may look to work as the human brain functions as detailed in [56]. The brain is in a continuous process to create the external world image in accordance with the data which it already knows and what it has already observed. This is done through a model which generates the signals hierarchically with an aim of reducing the errors in prediction via a bi-directional cascading of the cortical processing. Specifically, the brain selects, in terms of what it expects, a minute subset of signal(s) multitude which is sent by the retina. In this manner, majority of the signals which are generated in retina are not required to propagate via optical nerve as only those signals (observations) which deviate from the prediction are transmitted from the retina to the brain. In this manner, much energy is saved and this process could be replicated to the next generation semantic communications enabled systems.

### ii. Speech Communications

The incorporation of NLP within the speech communications will be beneficial in addition to including a step of speech recognition which will aid in the translation of speech to text. Further, to minimize the efforts of the forward error correction codes, automated word(s)/sentence(s) corrections can be included so that any error(s) at the bit/symbol level is compensated by the word(s)/sentence(s) technique. As an alternate, instead of re-transmission, speech bits which are lost due to interference/fading/crosstalk can be retrieved through context only.

Also, in regard to speech signal processing, an intelligent task will be the conversion of the speech signals to the corresponding text data which, in comparison to the usual automatic speech recognition technique, will account for speech signals characteristics. The process will include the mapping of every phoneme to a corresponding individual alphabet following which, there will be a requirement of concatenating all the alphabets to the corresponding sequence of words through a model of language which can be understood by all. Further, for this scenario, the semantic features which will be extracted will only contain the text characteristics whereas, other features are not transmitted by source. Hence, network traffic is considerably minimized ensuring no degradation in the network performance.

### iii. Video Communications

This use case expects significant improvements presently and in the future as it consumes much of the network resources. The open research problem is the manner in which the semantic aspect can be integrated with the existing video communications setup. A possible solution is to incorporate the meaning/context of the content(s) within the video. Specifically, using the previously existing frames of the video, an interpreter will be able to predict the current and future frames via an appropriately trained model for prediction.

Using the previous frames, interpreter is able to predict next frames via an appropriately trained model for prediction. The study in [57] has performed the video coding after the incorporation of a frame predicting method which is enabled by deep neural networks. In this study, when there occurs no significant change due to fading, the complete flow of the video is constricted with no changes at the semantic level. The interpreter at the receiver is able to reproduce a video, which may not be syntactically matching with the transmitted video; however, is a semantic equivalent. As a result, this technique generates major savings in regard to the transmitted power and/or the bandwidth.

Another concept in regard to the video communications is the process of segmentation of the semantics i.e., to implement machine vision tasks to segment all the objects within a scene so as to classify them based on a concept. The aforementioned is in turn advantageous since it will serve as a pre-processing step to further tasks such as, object detection, scene understanding, and scene parsing. In effect, semantic segmentation will analyse and classify the objects' nature and concept in addition to being able to recognize the objects and the corresponding shape within the scene. Therefore, semantics segmentation will be considered as one of the three basic steps of object detection, shape recognition and classification.

iv.  **Goal Oriented Communications**

In our view, the goal-oriented communications will be part of the opportunities which are offered by the Level 3 (effectiveness) layer. The success of this communication type will require the specification of a goal with immense clarity such that all the data is not transmitted; rather, only the most relevant data is sent by the transmitter resulting in optimized network performance and effective goal-oriented communication. Existing work on goal-oriented communication has focused on the concept of misunderstanding between the communicating parties arising due to the absence of an agreement on the protocol/language type to be used for communication [27, 28, 58]. However, we present a different view of goal-oriented communication in which the aim to communicate must be to fulfil a goal in which case the system performance will be dictated by the completion of the specific goal. In specific, effectiveness will have to evaluated based on the goal completion with the constraints on the resources. As an example, for the 6G networks, after the allocation of a chunk of the THz frequency, the aim will be to ensure that the transmitter and receiver send/receive the bits/symbols to complete the communication within the allocated frequency chunk only. Such an issue is addressed in [59-61]. The authors in [59] have addressed this issue for the case of edge learning in which the tools for learning are much closer to the user equipment to provision the applications with acute constraints of delay. The authors state that in such cases there will occur multiple trade-offs such as, between delay and consumption of power, between delay and accuracy, etc. In [60], the authors have addressed the trade-off between delay and accuracy considering an edge ML system enabled via an edge processor implementing the stochastic gradient descent (SGD) algorithm. Given every data packet's overhead and ration of computation and communicating rates, the authors have optimized size of the packet payload. The authors in [61] have proposed an algorithm which aims at maximizing the accuracy of learning under the constraints of delay. In [62], authors have proposed a distributed ML algorithm for edge operation of wireless equipment which aim to reduce an empirical function indicating loss via a remote server.

In view of addressing the aforementioned issue, we propose the following formulation. Considering our proposal, shown in Fig. 8, the novelty in our formulation exists in the feedback which is provided to the source encoder for goal oriented communication by the decision maker block. To elaborate, consider the goal of communication being either to conduct a classification or use parameters set $\xi$ comprised of an observation(s) set $m_i, 1,2,\ldots\ldots N$ for learning. Further, let the evaluation set be denoted as $M := \{m_i\}_{i=1}^{N}$. With the aforementioned, the aim will be to transmit the data, $D$, to a decision centre for making a decision, and the issue to be resolved will the manner in which the data must be encoded i.e., the mapping of $M$ to $D$ simultaneously provisioning the applications with highest accuracy and lowest power consumption. Further, initially, the data will have to be encoded at the transmitter for the removal of any redundancy simultaneously permitting ideal reconstruction or to reach a permissible compromise between the rate of encoding and distortion. In comparison to the aforementioned, we propose a strategy in which if communication occurs to complete a specific goal, then the source encoder must be designed such that it is able to achieve the accuracy which is desired over the classification simultaneously reducing the resources which are used or minimizing the decision time. Further, in comparison to reducing the errors while reconstructing the data, the source encoder will have to be tuned such that it falls in line with the learners' performance at the decision centre.

To state the problem formally, we assume that the parameter (random variable) to be estimated is denoted as $\phi$. The problem is analogous to searching a map of $D = f(M)$ such that there occurs a maximum compression of $D$. However, it must be noted that in this process of moving from $M$ to $D$ there should occur no loss of the data in regard to the recovery of $\xi$. In other words, the problem can be stated as: $I(M;\phi) = I(D;\phi)$ where $I(M;P)$ is the mutual information from $M$ to $Y$, and the corresponding solution is the reduced sufficient statistics of $M$. Further, given $m$, $f(m)$ is a sufficient statistics for $\phi$ and the joint pdf $p(m,\xi)$ is evaluated as [63]:

$$p(m;\xi) = g_\xi[f(m)]h(m). \tag{9}$$

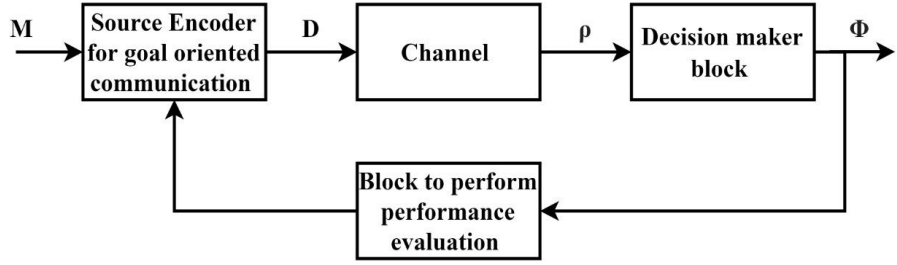

Fig. 8. Block diagram of goal-oriented communication.

However, there could be cases when multiple sufficient statistics exist. In this regard, considering the proposal in this study, it will be key to identify the minimal sufficient statistics. In specific, relative to $p(m;\xi)$, the sufficient statistics $f(m)$ will be minimal only if it is a function of every other statistic [61]. The aforementioned implies that if communication aim is to estimate parameter $\xi$ from data set $M$ then there will occur no data loss if instead of $M$, $f(M)$ is transmitted. This presents an advantage in terms of the entropy wherein, the entropy of $f(M)$ could be much lesser that that of $M$ which implies that encoding $f(M)$ may require much lesser bits in comparison to the number of bits required to encode $M$. In turn, this will aid in minimization of the transmitted bits simultaneously ensuring that there is no loss in the meaning (semantics) of the data.

Further, for cases when it is difficult to find the minimal sufficient statistics, we extend our proposal for general cases which follows the bottleneck principle as detailed in [64]. The formulation searches for an encoding function $D = f(M)$ which aims to achieve the best trade-off which exists between data loss in regard to estimation of $\phi$ which is obtained via the compression of $M$ and the bits amount required to encode $f(M)$. Mathematically, the generalized formulation is as follows [63]:

$$\min_{f(M)} I(M;D) - \gamma I(D;\phi). \qquad (10)$$

where $\gamma$ denotes a positive real number which assigns multiple weights to the two terms in the formulation. The data can be compressed to maximum value if the value of $\gamma$ is small; however, this compromises the accuracy of learning. To improve the accuracy of learning, the value of $\gamma$ must be large; however, this compromises the compression at the transmitter. The solution to the generalized problem can be achieved through the iterative data bottleneck technique which has been shown to be effective within the learning problems via the use of auto-encoders [64, 65]. Lastly, $\gamma$ can be adapted dynamically to evaluate best compromise between power consumption, latency, and learning accuracy. This will depend on both, the channel state and accuracy level which is achieved by the learner.

v. **Online Learning Communications**

In 6G networks, compared to the previous generation networks, machines will be contributing significantly via the enabling of AI at the network edge which will exist much closer to the end users [66, 67]. This in turn will ensure that (i) ML can be implemented for communication, computation, control and infrastructure in view of optimizing the network resource(s) usage, and (ii) communication can be implemented for ML with an aim to enable semantic aware latency critical services via the ML algorithms' distributed implementation. For supporting the aforementioned, 6G networks design will have to include learning tools so as to tune the network in response to changes in the requirements and the constraints [68]. In addition, we also envision that the inductive ML techniques will be integrated with the deductive semantic aspects to achieve the further development of (i) ML algorithms through semantic communication for exploiting the context aspects for improving the learning capabilities, using the semantic data efficiently, and enhancing the robust nature of the system in response to adverse attacks, and (ii) semantic communications through ML techniques to infer the context efficiently so as to improve the system efficiency.

vi. **Intelligent Communications**

6G wireless networks will provision next-generation services which will require the system to take effective decisions under the constraints of stringent latency and jitter bounds. In view of the aforementioned, edge computing has been identified as an emerging technology which will ensure that the computations will occur at the network edge instead of a distant cloud server. This will in turn ensure that the end users will be able to access all the computations and resources incurring very low delays. In general, latency occurs due to communication delay, computation delay, and infrastructure (resource(s)) access delay. In addition, if the procedures which are latency critical also need to be controlled, then a control delay also occurs. Hence, for the 6G networks, a joint inclusion of communication, computation, infrastructure, and control must be considered for the design [69]. Such a proposal is presented by the authors in [70, 71] considering case of cellular edge computing and further, in [72], authors have proposed an edge controller to implement a strategy which aims at optimized controlling of the systems which enable advanced driving and controlling. However, in all these studies the authors have not considered the latency and power consumption by the data which is to be transmitted/received and processed. In this regard, the authors in [73] have considered the joint optimization of the resources relevant for communication and computation considering that the offloading of these resources will result in latency. Hence, the authors have proposed a setup which is multi-user i.e., one edge computing host is deployed for serving many small cells. Also, in [74, 75], the authors have proposed an algorithm which aims to optimizing communication and computation resources together in a dynamic environment. As an extension, the authors in [76] have proposed the use of a dynamic convex optimization technique in cases when the optimizer is not aware of the exact system state. Lastly, in addition to joint consideration of communication and computation, infrastructure and communication must also be considered together. This is mainly required in cases wherein; the applications demand the contents dynamically to ensure that the desired latency constraints are satisfied.

Further, to deliver the contents in an intelligent manner at the edge, dynamic access of the infrastructure will be key in view of minimizing the request initialization and the corresponding content delivery. In the networks which allow for infrastructure access at the edge, majority of the content will be present at the edge for access thereby, resulting in the ubiquitous visibility of the data at the user equipment. In specific, the infrastructure access rules will be a key enabler if it allows for the dynamic data storage in response to the demands estimated through an efficient prediction algorithm using multiple variables. Further, the variables will include the infrastructure deployment, content storage, and content routing [77].

In regard to infrastructure, thus far, the main issue has been the formulation of policies for moving the relevant content throughout the network. However, in the semantic communications, we envision that, in addition to only moving the relevant contents, it will also be required to move the base of knowledge systems and machines for dynamic provisioning of the demands much closer to the user equipment. The aforementioned has received much research attention as it poses the challenging issue in regard to the computation and infrastructure access for applications which are latency sensitive. Specifically, when contents are distributed within the network, it results in multiple issues in terms of bandwidth, latency, privacy, security, etc. As a solution, the coded distributed computing may be employed which resorts to the use of coding theory to introduce structured computation with redundancy [78].

vii. Machine Learning for Wireless Networks

Generally, when the supervised learning type of ML technique is implemented, there is trade-off between requirement of large labelled data amount and the achievement of the desired high performance. This implies that there is immense manual intervention with the supervised learning in view of providing the labelled data/examples. In comparison, the unsupervised learning type of ML technique does not require any labelled data/examples, and the learner aims to search for specific patterns within the data. In addition, few applications also implement the graph enabled approaches which aim at capturing the pair-wise relations among the variables; however, this is not efficient as it does not extract the complete information.

For the effective operation of systems which are enabled by wireless network, mechanisms must be proactive which will need efficient ML prediction methods to operate at the network edge. Over the past decade, deep neural networks (DNNs), a class of supervised learning, have been demonstrated to provide higher performance in comparison to other techniques, especially for applications such as, recognition of the sound and classification of the image [79]. Recently, the aspect of explainability has gained much importance owing to the widespread use of DNN, which find key applications in multiple fields. However, the manner in which an input of DNN generates output is not often evident. In specific, even when clarity exists in relation between input and output, final output of network weights, which result from training phase involving highly non-linear optimization, does not match with the theoretical guarantees. A key issue to be addressed in the explainable ML technique is whether the algorithm provides information which permits the user to relate the characteristics of the input features with the corresponding output. Also, the convolutional neural networks (CNN) have also shown suitability in regard to the sound and images owing to the technique's ability to intrinsically sparsify the edges amount from one layer to the next layer. As an extension, studies such as [80] have addressed the combination of graph-enabled representation with the learning aspect in view of operating with the data residing on the graphs rather than on the regular grids.

In specific, the application of supervised learning, in which the testing and learning phases have a clear distinction, to any communication network provides immense opportunity for training the learner through aggregated models which are stochastic in nature via a simulator which can generate the input data, the channels, and the output [81, 82]. Thus, the learner learns through large number of labelled examples. However, for the wireless networks with varying channel over time, dynamic (online) learning techniques are better suited as they have a combined testing and learning phase which updates with time. In this regard, the reinforcement Learning (RL) methods and the random optimization methods may be implemented. In RL, learning of an agent occurs through the observation of its own action(s) without the assumption of any a-priori observation model [83]. In the random optimization technique, an adaptive online process is implemented which updates a per the actions and the knowledge which is available at that instant regarding the variables which are involved [84]. As an extension, an advance is demonstrated in [85] by incorporating the multi-way relations via topological signal processing (TSP) which is helpful when dealing with observations which are associated to graph edges. Further, the ML methods are applicable to physical layer of the network via auto-encoder (AE) [86, 87], a recurrent neural network (RNN) [88], and a generative adversarial network (GAN) [89], and it can also be extended to higher layers through a complex AI framework as detailed in [90].

In the 6G wireless networks, there will exist tools for learning pervasively at the network edges which will pose numerous challenges. Further, within the network, multiple user equipment will generate large data which will require collaborative learning for the analysis in view of improving the learning technique's performance. These collaborative learning methods will demand data exchange which will in turn raise serious privacy/security issues which need to be addressed. As a solution, federated learning (FL) can be implemented in which the model parameter(s) learning occurs at a central unit/data centre/edge host, and the data is stored at the peripheral nodes [91, 92]. In the FL of centralized type, rather than sending the data to a remote server, the user equipment share parameters' local estimates which are to be learned. Therefore, the privacy issue is resolved and the user equipment performance is also improved.

Considering the aforementioned, a typical FL enabled network will have the following objective function:

$$\min_{m} \sum_{i=1}^{N} p_i f_i(d_i; m). \qquad (11)$$

where, the user equipment's ($i$) empirical loss function is denoted by $f_i(d_i; m)$, the parameter vector which is to be learnt is denoted as $m$, and $p_i$ denotes the weighting coefficient which highlights the data importance which has been collected by user $i$ within $m's$ estimation. Also, $p_i$ is defined as $p_i \geq 0$ and $\sum_{i=1}^{N} p_i = 1$. Further, the weights $p_i$ can be selected as

$$p_i = \frac{n_i}{\sum_{i=1}^{N} n_i}. \qquad (12)$$

where $n_i$ denotes the amount of examples which are observed by the $i^{th}$ device.

It must be noted that FL follows an iterative technique wherein, for each iteration $n$, as opposed to transmitting the local data $m_i$, every user equipment transmits the local estimate $\hat{m}_i[n]$ to a hub which in turn transmits back an update factor which accounts for the data which is received via all the cooperating nodes. This objective function converges to a global optimum under the conditions of convexity [92]. Also, this setting is simple; however, it faces the major challenges owing to communication channel heterogeneity, and behaviour of the local user equipment' model. Therefore, an extension technique multi-task FL can be implemented [93, 94].

From the aforementioned, it is clear that the ML technique is driven through data which aids in learning which is mostly inductive. However, human learning is deductive wherein, accumulation of experience and knowledge over a period of time helps in the continuous updation of the base of knowledge. Further, it is known that the inductive learning models such as, DL enabled systems result in ambiguity(s) as they are always searching for correlations within the data whereas, search for semantics (meaning) has much more value [95].

In this context, we believe that ML techniques will take significant leaps forward by incorporating the knowledge from external world and including the context within the decision making process. Further, the 6G networks will facilitate the merging of ML techniques and systems which are enabled knowledge representation [96, 97]. As a novelty, semantic communication will ensure that a base of knowledge exists at all the network nodes so as to enable the interpretation of semantics. This will require new ML techniques which will include the schemes for representation and reasoning in effect offering multiple capabilities to effectively optimize the network resource(s) usage in view of semantic and goal-oriented communications. Overall, the new ML techniques are envisioned to be more efficient and reliable owing to the inclusion of transfer learning and knowledge sharing mechanisms.

## V. Issues and Challenges

In this section, we present the various challenges and issues which exist in the research over semantic communications and networks.

1. Immersive XR is a combination of AR, VR, and mixed reality (MR) which includes the advantages of the three methods. XR technology will integrate with wireless networking, edge computing, and AI/ML for offering complete immersive experiences to the humans for multiple applications. The designs for XR are similar to the ones required for AR/VR; however, there will be stringent requirements in regard to the accuracy and the sensing human characteristics diversity, data rates, and latency [98]. In addition, large scale AI/ML models will be required to be implemented in view of efficient training and inference support.

2. Holographic Communication will involve the transmission and reception of 3D holograms of human or physical objects [99]. Depending on the high resolution, wearables, and AI/ML, user equipment will render the 3D holograms for displaying the remote users or machines local presence, to create increased realistic local presence of a remote human or object. This will require an enhancement in the users' visual perception for improving the virtual interaction efficiency. In this regard, haptic information's high-resolution encoding, colours, positions, and human/object tilts will be highly desired which will need extremely high data rates, and stringent delay constraints [100]. Hence, it will be crucial to boost semantic communication's processes viz., encoding, decoding, transmission, reception, etc., efficiency and speed to higher levels such that the holographic communications will be able to satisfy the main aim of real-time and seamless integration of human-to-machine and machine-to-machine techniques.

3. All Sense Communication will be included within the 6G communications via a sensors ensemble which will be a wearable or will be mounted over every device [101]. In combination with holographic communications, all sense data will be effectively integrated for realizing a close-to-real feelings of the remote environments thereby, facilitating the tactile communications and haptic control [102]. The challenge will be to provision the exponential complexity within the applications which will arise due to diversified types of sensing signals which will create multiple new data dimensions.

4. In comparison to the sampling and reconstruction method used for syntactic communication, in semantic communications compressed sensing will be implemented for performing specific signal processing methods which will occur directly in compressed domain without complete signal reconstruction [103]. Therefore, a major challenge will be to develop an optimal sampling technique (theory) which will combine signal sparsity and semantics in view of real-time prediction and/or reconstruction considering the various constraints and latency. For achieving the optimal trade-offs between the metrics used for semantic communications and the network energy consumption, major revision is required in the feedback structure, packetization, scheduling and resource assignment. This further implies that there is a need to abandon simplistic assumptions and simple metrics, and in favour of the advanced ML techniques.

5. In view of savings on miscellaneous operations over packet data, the optimization of packet structure i.e., a clear differentiation between data and miscellaneous data is required. Also, keeping the aims of semantic communications in mind, the following require further research (i) the role which feedback will play considering real time channel environment, (ii) the retransmissions over the link layer under real channel conditions, and (iii) the error rates which can be achieved considering lengths of the block which are non-asymptotic in nature.

6. Another key challenge is related to scheduling and resource allocation policies for semantic communications which will also consider the optimization of energy within the system. Further, advanced access techniques, flow control mechanisms, and much broader metric for semantic communications remains an open research problem which requires timely solutions. Specifically, in regard to the metrics, it must also be noted that considering the diversity of humans, new composite metrics need to be developed so as to cover the key experiences.

7. For a semantic enabled network, the complexity will be high as the network will have to provision a system in which the knowledge is shared very close to all the users [104]. This will require the development of a framework which is mathematically enabled, and which will aid in the evaluation of performance limits imposed by the semantic network. Further, tracking and modelling the upgrade in human knowledge to continuously update the base of knowledge will aid in the improvement of communication efficiency, and the minimization of error probability within the semantic network. This aspect requires further investigation.

Finally, in Table 2, we suggest the various directions which can be adopted in response to the existing research challenges on semantic communications.

Table 2. A summary of various existing challenges in research on semantic communications and possible directions to obtain solutions.

| Sl. No. | Existing Challenges | Proposed Directions |
|---|---|---|
| 1 | Holographic Communications | Provide enhancement in encoding, decoding, transmission, reception, etc., to increase the efficiency and speed to higher levels. Develop a model which ensure that holographic communications satisfy real-time and seamless integration of human-to-machine and machine-to-machine techniques. |
| 2 | XR | Satisfy stringent requirements in regard to accuracy and sensing human characteristics diversity, data rates, and latency. Implement large scale AI/ML models for efficient training and inference support. |
| 3 | All Sense Communications | Develop models to provision the exponential complexity within applications arising due to varied sensing signals types which will create multiple new data dimensions. |
| 4 | Sampling and Reconstruction | Develop an optimal sampling technique to combine signal sparsity and semantics for real-time prediction and/or reconstruction considering various constraints and latency. |
| 5 | Metrics | Develop advances metrics in favour of the advanced ML techniques. Examples: freshness, relevance, value, etc. [29]. |
| 6 | Semantic Network | Develop a mathematical framework which will aid in performance evaluation limits imposed by the semantic network. |
| 7 | Scheduling and resource allocation policies | Formulate new random/scheduled access policies in regard to broader semantic metrics. |
| 8 | Knowledge modelling and tracking | Mechanism to continuously update the base of knowledge automatically for improving communication efficiency, and reduce error probability within the semantic network. |
| 9 | Multi-objective Stochastic Optimization | To develop a model which will solve the multiple criteria optimization with goal-oriented, end-user perceived utilities to enable relative priority degree among various attributes of information. |
| 10 | Flow control techniques | To develop optimal operating protocols for higher layer operation to improve the network resource(s) usage. |

# VI. Conclusion

In this article, we have surveyed the most recent advances which will aid the enabling of semantic communications for intelligent wireless networks. We reviewed the concept of semantic communications and the application to intelligent wireless networks, which presents a paradigm deviation from the conventional method of following the Shannon's information theory for communication. From the survey, we gathered that the semantic method of communication offers a potential to generate enormous benefits, increased effectiveness and reliability in addition to presenting a goal-oriented manner of information exchange. Further, the aforementioned is possible without the need to allocate increased resources (bandwidth, energy, etc.), and requires only the correct identification of the appropriate data. This implies that the semantic communication enabled wireless networks will only require to extract the correct meaning/context of the transmitted data, and use the same to decode the received data to ensure that goal-oriented communication occurs. This in turn focuses on the sustainability of the wireless networks and breaks away from the convention of providing more resources for provisioning advanced applications. Conversely, semantic communication aims to utilize untapped capability(s) such as, communication, computation, infrastructure and control of a communication system to represent the knowledge so as to communicate only the relevant data between the parties which interact with one another. In regard to the challenges, we found that semantic communications will require much research in regard to the (i) distributed computing techniques' implementation so as to learn, and then extract the precise context from the received data, (ii) exploitation of the appropriate system(s) to represent the knowledge, and (iii) identification and exploitation of the most relevant data in view of goal oriented communications. Once implemented, such a new framework will enable a form of learning which will be gather enhanced benefits by the introduced aspects of semantics thereby, migrating from a completely inductive technique to a strategy which will encompass an interplay of both, inductive and deductive techniques. Further, this will also enable a learning mechanism which, in addition to learning through examples, will construct abstract models to guide further learning. Lastly, we have also presented future directions to the multiple existing challenges with a hope that this will spur further research on semantic communications for next generation intelligent wireless networks.

# Declarations

**Funding**
The authors declare that no funds, grants, or other support were received during the preparation of this manuscript.

**Conflict of Interests**
The authors have no conflict of interests to disclose.

**Data Availability**
Data sharing not applicable to this article as no datasets were generated or analysed during the current study.

The pre-print version of this article is available at: http://arxiv.org/abs/2202.03705

**Code Availability**
Not Applicable

**Author Contributions**
All authors contributed to the study conception and design. All authors read and approved the final manuscript."

# References

1. M. Shafi, A. F. Molisch, P. J. Smith, T. Haustein, P. Zhu, P. D. Silva, F. Tufvesson, A. Benjebbour, and G. Wunder, "5G: A tutorial overview of standards, trials, challenges, deployment, and practice", *IEEE J SeL Area Comm.*, vol. 35, no. 6, pp. 1201-1221, 2017.
2. K. David and H. Berndt, "6G Vision and Requirements: Is There Any Need for Beyond 5G?", *IEEE Veh. Technol. Mag.*, vol. 13, pp. 72–80, 2018.
3. 'The Vision of 6G: Bring the next hyper-connected experience to every corner of life', Samsung, White Paper, 2020. [Online]. Available: https://news.samsung.com/global/samsungs-6g-white-paper-lays-out-the-companys-vision-for-the-next-generation-of-communications-technology
4. N. Rajatheva *et al.*, "Scoring the Terabit/s Goal: Broadband Connectivity in 6G", 2021. [Online]. Available: https://arxiv.org/abs/2008.07220
5. H. Saarnisaari *et al.*, 6G White Paper on Connectivity for Remote Areas [White paper]. (6G Research Visions, No. 5). University of Oulu. 2020. [Online]. Available: http://urn.fi/urn:isbn:9789526226750
6. N. Rajatheva *et al.*, "White paper on broadband connectivity in 6G," June 2020. [Online]. Available: http://urn.fi/urn:isbn:9789526226798
7. K. Tekbıyık, A. R. Ekti, G. K. Kurt, and A. Görçinad, "Terahertz band communication systems: Challenges, novelties and standardization efforts", *Physical Communication*, vol. 35, 2019.
8. C. E. Shannon, A mathematical theory of communication, *The Bell system technical journal*, vol. 27, pp. 379-423, 1948.
9. H. Sarieddeen *et al.*, Next Generation Terahertz Communications: A Rendezvous of Sensing, Imaging, and Localization, *IEEE Commun. Mag.*, vol. 58, no. 5, pp. 69-75, 2020.
10. Y. Corre, G. Gougeon, J.-B. Dor´e, S. Bicaıs, B. Miscopein, E. Faussurier, M. Saad, J. Palicot, and F. Bader, "Sub-thz spectrum as enabler for 6G wireless communications up to 1 tbit/s", 6G Wireless Summit, Levi Lapland, Finland. hal-01993187, 2019.
11. P. Popovski, O. Simeone, F. Boccardi, D. Gunduz, and O. Sahin, "Semantic effectiveness filtering and control for post-5G wireless connectivity", *Journal of the Indian Institute of Science*, vol. 100, pp. 435-443, 2020.
12. M. H. Alsharif, A. H. Kelechi, M. A. Albreem, S. A. Chaudhry, M. S. Zia, and S. Kim, "Sixth Generation (6G) Wireless Networks: Vision, Research Activities, Challenges and Potential Solutions", *Symmetry, MDPI*, vol. 12, no. 4, pp. 676, 2020.
13. J. Kohlas, "Information algebras: Generic structures for inference", Springer Science & Business Media, 2012.
14. C. S. Calude, "Information and randomness: an algorithmic perspective", Springer Science & Business Media, 2013.
15. B. Guler, A. Yener, and A. Swami, "The semantic communication game", *IEEE Transactions on Cognitive Communications and Networking*, vol. 4, no. 4, pp. 787-802, 2018.
16. M. Kountouris, and N. Pappas, "Semantics-empowered communication for networked intelligent systems", arXiv preprint, 2020. [Online]. Available: https://arxiv.org/abs/2007.11579
17. H. Xie, Z. Qin, G. Y. Li, and B. H. Juang, "Deep learning enabled semantic communication systems", arXiv preprint, 2020. [Online]. Available: https://arxiv.org/abs/2006.10685
18. Z. Weng, and Z. Qin, 'Semantic Communication Systems for Speech Transmission', *IEEE Journal on Selected Areas in Communications*, vol. 39, no. 8, pp. 2434–2444, 2021.
19. H. Tong, Z. Yang, S. Wang, Y. Hu, O. Semiari, W. Saad, and C. Yin, "Federated Learning for Audio Semantic Communication", Front. Comms. Net., 2021. [Online]. Available: https://www.frontiersin.org/articles/10.3389/frcmn.2021.734402/full '

20. P. Zhang, W. Xu. H. Gao, K. Niu, X. Xu, X. Qin, C. Yuan, Z. Qin, H. Zhao, J. Wei, and F. Zhang, "Toward Wisdom-Evolutionary and Primitive-Concise 6G: A New Paradigm of Semantic Communication Networks", *Engineering, Elsevier,* 2021. [Online]. Available: https://www.sciencedirect.com/science/article/pii/S2095809921004513
21. G. Shi, Y. Xiao, Y. Li, D. Gao, X. Xie, "Semantic communication networking for the intelligence of everything", *Chinese Journal on Internet of Things*, vol. 5, no. 2, pp. 26-36, 2021.
22. M. Sana, and E. C. Strinati, "Learning Semantics: An Opportunity for Effective 6G Communications", [Online]. Available: https://arxiv.org/abs/2110.08049
23. G. Shi, Y. Xiao, Y. Li, and X. Xie, "From Semantic Communication to Semantic-Aware Networking: Model, Architecture, and Open Problems", *IEEE Communications Magazine*, 2021. [Online]. Available: https://arxiv.org/abs/2012.15405
24. P. Zhang, L. Li, K. Niu, Y. Li, G. Lu, and Z. Wang, "An intelligent wireless transmission toward 6G", *Intelligent and Converged Networks,* vol. 2, no. 3, pp. 244−257, 2021.
25. J. Bao, P. Basu, M. Dean, C. Partridge, A. Swami, W. Leland, and J. A. Hendler, "Towards a theory of semantic communication", in 2011 IEEE Network Science Workshop, IEEE, 2011, pp. 110-117.
26. P. Basu, J. Bao, M. Dean, and J. Hendler, "Preserving quality of information by using semantic relationships", Pervasive and Mobile Computing, vol. 11, pp. 188-202, 2014.
27. B. Juba, and M. Sudan, "Universal semantic communication ii: A theory of goal-oriented communication", in: Electronic Colloquium on Computational Complexity (ECCC), vol. 15, 2008.
28. O. Goldreich, B. Juba, and M. Sudan, "A theory of goal-oriented communication", Journal of the ACM (JACM), vol. 59, pp. 1-65, 2012.
29. E. Uysal, *et. al.,* "Semantic Communications in Networked Systems", 2021. [Online]. Available: https://arxiv.org/abs/2103.05391
30. G. Shi, D. Gao, X. Song, J. Chai, M. Yang, X. Xie, L. Li, and X. Li, "A new communication paradigm: from bit accuracy to semantic fidelity," 2021. [Online]. Available: https://arxiv.org/abs/2101.12649
31. S. J. Russell, and P. Norvig, "'Artificial intelligence-a modern approach", 4th edition, 2020.
32. X. Han, and C. K. Kwoh, Natural Language Processing Approaches in Bioinformatics", in Encyclopedia of Bioinformatics and Computational Biology, 2019.
33. S. Shakkottai, T.S. Rappaport, and P.C. Karlsson, "Cross-Layer Design for Wireless Networks," IEEE Commun. Mag., vol. 41, no. 10, pp. 74–80, 2003.
34. B. Fu, Y. Xiao, H. J. Deng and H. Zeng, "A Survey of Cross-Layer Designs in Wireless Networks," in IEEE Communications Surveys & Tutorials, vol. 16, no. 1, pp. 110-126, 2014.
35. J. Bing, S. Liu, and Y. Yang, "Fractal cross-layer service with integration and interaction in internet of things", International Journal of Distributed Sensor Networks, vol. 10, no. 3, pp. pp. 760, 2014.
36. H. Berndt, "Towards 4G Technologies: Services with Initiative", John Wiley & Sons, New York, NY, USA, 2008.
37. Y. H. Liu, "Introduction to Internet of Things", Science Press, Beijing, China, 2010.
38. L. Atzori, A. Iera, and G. Morabito, "The Internet of Things: a survey," Computer Networks, vol. 54, no. 15, pp. 2787–2805, 2010.
39. S. Reddy, V. Samanta, J. Burke, D. Estrin, M. Hansen, and M. Srivastava, "Mobisense—mobile network services for coordinated participatory sensing," in Proceedings of the International Symposium on Autonomous Decentralized Systems (ISADS '09), pp. 231–236, Athens, Ga, USA, 2009.
40. M. Achir, A. Abdellia, L. Mokdad, and J. Benothman, "Service discovery and selection in IoT: A survey and a taxonomy", *Journal of Network and Computer Applications, Elsevier*, 2022. https://doi.org/10.1016/j.jnca.2021.103331
41. A. H. Ahmed, N. M. Omar, and H. M. Ibrahim, "Secured Service Discovery Technique in IoT", Journal of Communications, vol. 14, no. 1, pp. 40-46, 2019

42. L. Atzori, A. Iera, and G.Morabito, "The Internet of Things: a survey," Computer Networks, vol. 54, no. 15, pp. 2787–2805, 2010.
43. B. Jia, "Research on Semantic-Based Service Architecture and Key Algorithms for the Internet of Things", Jilin University, 2013, (Chinese).
44. A. Gangemi, P. Mika, M. Sabou, and D. Oberle, "Technical Report: An Ontology of Services and Service Descriptions, Technical Report, 2003. [Online]. Available: https://www.researchgate.net/publication/228938556_Technical_Report_An_Ontology_of_Services_and_Service_Descriptions
45. J. Ballé, A. Shrivastava, and G. Toderici, "End-to-end Learning of Compressible Features", In IEEE International Conference on Image Processing (ICIP). IEEE, pp. 3349–3353, 2020.
46. Tu, Hanyue, *et al*. "Semantic Scalable Image Compression with Cross-Layer Priors.", in Proceedings of the 29th ACM International Conference on Multimedia, 2021.
47. S. Dodge and L. Karam, "Understanding how image quality affects deep neural networks", In 8th international conference on quality of multimedia experience (QoMEX). IEEE, pp. 1–6, 2016.
48. K. He, X. Zhang, S. Ren, and J. Sun, "Deep residual learning for image recognition", in Proceedings of the IEEE conference on computer vision and pattern recognition, pp. 770–778, 2016.
49. G. Huang, Z. Liu, L. Van Der Maaten, and K. Q Weinberger, "Densely connected convolutional networks", in Proceedings of the IEEE conference on computer vision and pattern recognition, pp. 4700–4708, 2017.
50. K. Simonyan and A. Zisserman, "Very deep convolutional networks for large-scale image recognition", 2014. [Online]. Available: arXiv:1409.1556.
51. A. Dosovitskiy and T. Brox, "Inverting visual representations with convolutional networks", in Proceedings of the IEEE conference on computer vision and pattern recognition, pp. 4829–4837, 2016.
52. Z. Chen, K. Fan, S. Wang, L. Duan, W. Lin, and A. C. Kot, "Toward intelligent sensing: Intermediate deep feature compression", *IEEE Transactions on Image Processing*, vol. 29, pp. 2230–2243, 2019.
53. Z. Chen, K. Fan, S. Wang, L. Yu Duan, W. Lin, and A. Kot, "Lossy intermediate deep learning feature compression and evaluation", on Proceedings of the 27th ACM International Conference on Multimedia, pp. 2414–2422, 2019.
54. H. Choi and I. V Bajić, "Deep feature compression for collaborative object detection", in Proceedings 25th IEEE International Conference on Image Processing (ICIP). IEEE, pp. 3743–3747, 2018.
55. H. Choi and I. V Bajić, "Near-lossless deep feature compression for collaborative intelligence", in Proceedings IEEE 20th International Workshop on Multimedia Signal Processing (MMSP). IEEE, pp. 1–6, 2018.
56. A. Clark, "Whatever next? predictive brains, situated agents, and the future of cognitive science", *Behavioral and brain sciences*, vol. 36, pp. 181-204, 2013.
57. H. Choi, and I. V. Bajic, "Deep frame prediction for video coding", *IEEE Transactions on Circuits and Systems for Video Technology*, vol. 30, no. 7, pp. 1843-1855, 2019.
58. B. Juba, "Universal semantic communication", Springer Science & Business Media, 2011.
59. J. Park, S. Samarakoon, M. Bennis, and M. Debbah, "Wireless network intelligence at the edge", *Proceedings of the IEEE*, vol. 107, pp. 2204-2239, 2019.
60. N. Skatchkovsky, and O. Simeone, "Optimizing pipelined computation and communication for latency-constrained edge learning", *IEEE Communications Letters*, vol. 23, pp. 1542-1546, 2019.
61. U. Mohammad, and S. Sorour, "Adaptive task allocation for mobile edge learning", 2019. [Online]. Available: https://arxiv.org/abs/1811.03748
62. M. M. Amiri, and D. Gunduz, "Machine learning at the wireless edge: Distributed stochastic gradient descent over-the-air", 2019. [Online]. Available: https://arxiv.org/abs/1901.00844
63. E.C. Strinati, and S. Barbarossa, "6G networks: Beyond Shannon towards semantic and goal-oriented communications, *Computer Networks, Elsevier*, vol. 190, 2021.

64. N. Tishby, F. C. Pereira, and W. Bialek, "The information bottleneck method", 2000. [Online]. Available: https://arxiv.org/abs/physics/0004057
65. O. Shamir, S. Sabato, and N. Tishby, "Learning and generalization with the information bottleneck", *Theoretical Computer Science*, vol. 411, pp. pp. 2696-2711, 2010.
66. Z. Zhou, X. Chen, E. Li, L. Zeng, K. Luo, and J. Zhang, Edge intelligence: Paving the last mile of arti_cial intelligence with edge computing, *Proceedings of the IEEE*, vol. 107, pp. 1738-1762, 2019.
67. E. Peltonen *et al.,* "6G white paper on edge intelligence", 2020. [Online]. Available: https://arxiv.org/abs/2004.14850
68. A. Chaoub *et al.,* "6G for Bridging the Digital Divide: Wireless Connectivity to Remote Areas," *IEEE Wireless Communications*, pp. 1-9, 2021. [Online]. Available: https://arxiv.org/abs/2009.04175
69. S. Barbarossa, S. Sardellitti, E. Ceci, and M. Merluzzi, "The edge cloud: A holistic view of communication, computation, and caching", *Cooperative and Graph Signal Processing, Elsevier*, pp. 419-444, 2018. https://doi.org/10.1016/B978-0-12-813677-5.00016-X
70. A. Ndikumana, N. H. Tran, T. M. Ho, Z. Han, W. Saad, D. Niyato, and C. S. Hong, "Joint communication, computation, caching, and control in big data multi-access edge computing", *IEEE Transactions on Mobile Computing*, vol. 19, pp. 1359-1374, 2020.
71. N. Anselme, "Intelligent Edge: Joint Communication, Computation, Caching, and Control in Collaborative Multi-access Edge Computing", Ph.D. thesis, Kyung Hee University, South Korea, 2019.
72. Z. Wang, Y. Gao, C. Fang, Y. Sun, and P. Si, "Optimal control design for connected cruise control with edge computing, caching, and control", in: IEEE INFOCOM-2019) IEEE Conference on Computer Communications Workshops (INFOCOM WKSHPS)), pp. 1-6, 2019.
73. S. Sardellitti, G. Scutari, and S. Barbarossa, "Joint optimization of radio and computational resources for multicell mobile-edge computing", *IEEE Transactions on Signal and Information Processing over Networks*, vol. 1, pp. 89-103, 2015.
74. Y. Mao, J. Zhang, S. Song, and K. B. Letaief, "Stochastic joint radio and computational resource management for multi-user mobile-edge computing systems", *IEEE Transactions on Wireless Communications*, vol. 16, pp. 5994-6009, 2017.
75. M. Merluzzi, P. Di Lorenzo, S. Barbarossa, and V. Frascolla, "Dynamic computation o_oading in multi-access edge computing via ultra-reliable and low-latency communications", *IEEE Transactions on Signal and Information Processing over Networks*, vol. 6, pp. 342-356, 2020.
76. T. Chen, S. Barbarossa, X. Wang, G. B. Giannakis, and Z. L. Zhang, "Learning and management for internet of things: Accounting for Adaptivity and scalability", *Proceedings of the IEEE*, vol. 107, pp. 778-796, 2019.
77. G. Paschos, G. Iosifdis, and G. Caire, "Cache optimization models and algorithms", 2020. [Online]. Available: https://arxiv.org/abs/1912.12339
78. S. Li, and S. Avestimehr, "Coded computing: Mitigating fundamental bottlenecks in large-scale distributed computing and machine learning", Now Foundations and Trends, 2020. [Online]. Available: https://www.nowpublishers.com/article/DownloadSummary/CIT-103
79. Y. LeCun, Y. Bengio, and G. Hinton, "Deep learning", *Nature*, vol. 521, pp. 436-444, 2015.
80. M. M. Bronstein, J. Bruna, Y. LeCun, A. Szlam, and P. Vandergheynst, "Geometric deep learning: going beyond euclidean data", *IEEE Signal Processing Magazine*, vol. 34, pp. 18-42, 2017.
81. I. H. Sarker, "Machine Learning: Algorithms, Real-World Applications and Research Directions", *SN COMPUT. SCI.*, vol. 2, pp. 160, 2021. https://doi.org/10.1007/s42979-021-00592-x
82. A. Telikani, A. Tahmassebi, W. Banzhaf, and A. H. Gandomi, "Evolutionary Machine Learning: A Survey", *ACM Comput. Surv.,* vol. 54, no. 8, 2021.
83. R. Kirk, A. Zhang, E. Grefenstette, and T. Rocktäschel, "A Survey of Generalisation in Deep Reinforcement Learning", 2021. [Online]. Available: https://arxiv.org/abs/2111.09794

84. S. Sun, Z. Cao, H. Zhu and J. Zhao, "A Survey of Optimization Methods From a Machine Learning Perspective," in *IEEE Transactions on Cybernetics*, vol. 50, no. 8, pp. 3668-3681, 2020.
85. S. Barbarossa, and S. Sardellitti, "Topological signal processing over simplicial complexes", *IEEE Transactions on Signal Processing,* vol. 68, pp. 2992-3007, 2020.
86. S. Dorner, S. Cammerer, J. Hoydis, and S. Ten Brink, "Deep learning based communication over the air", *IEEE Journal of Selected Topics in Signal Processing,* vol. 12, pp. 132-143, 2017.
87. E. Balevi, and J. G. Andrews, "One-bit ofdm receivers via deep learning", *IEEE Transactions on Communications,* vol. 67, pp. 4326-4336, 2019.
88. N. Farsad, and A. Goldsmith, "Neural network detection of data sequences in communication systems", *IEEE Transactions on Signal Processing*, vol. 66, pp. 5663-5678, 2018.
89. H. Ye, G. Y. Li, B. H. F. Juang, and K. Sivanesan, "Channel agnostic end-to- end learning based communication systems with conditional gan", in 2018 IEEE Globecom Workshops (GC Wkshps), IEEE, pp. 1-5, 2018.
90. G. Dandachi, A. De Domenico, D. T. Hoang, and D. Niyato, "An artificial intelligence framework for slice deployment and orchestration in 5G networks", IEEE Transactions on Cognitive Communications and Networking, vol. 6, no. 2, pp. 858-871, 2020.
91. Q. Yang, Y. Liu, T. Chen, and Y. Tong, "Federated machine learning: Concept and applications", *ACM Transactions on Intelligent Systems and Technology (TIST)*, vol. 10, pp. 1-19, 2019.
92. T. Li, A. K. Sahu, A. Talwalkar, and V. Smith, "Federated learning: Challenges, methods, and future directions", *IEEE Signal Processing Magazine*, vo. 37, pp. 50-60, 2020.
93. V. Smith, C. K. Chiang, M. Sanjabi, and A. S. Talwalkar, "Federated multitask learning", *Advances in Neural Information Processing Systems*, pp. 4424-4434, 2017.
94. D. Shome, O. Waqar, W. U. Khan, "Federated learning and next generation wireless communications: A survey on bidirectional relationship", 2021. [Online]. Available: https://arxiv.org/abs/2110.07649
95. M. Abdar *et. al.,* "A review of uncertainty quantification in deep learning: Techniques, applications and challenges", *Information Fusion, Elsevier*, vol. 76, pp. 243-297, 2021.
96. H. Kim, "Artificial Intelligence for 6G", *Springer International Publishing*, 2022.
97. V. P. Rekkas, S. Sotiroudis, P. Sarigiannidis, S. Wan, G. K. Karagiannidis, and S. K. Goudos, "Machine Learning in Beyond 5G/6G Networks—State-of-the-Art and Future Trends, *Electronics, MDPI*, vol. 10, pp. 2786, 2021.
98. S. Rokhsaritalemi, A. Sadeghi-Niaraki, and Soo-Mi Choi, "A Review on Mixed Reality: Current Trends, Challenges and Prospects", *Appl. Sci., MDPI*, vol. 10, pp. 636, 2020. doi:10.3390/app10020636
99. S. A. Shoydin, and A. L. Pazoev, "Transmission of 3D Holographic Information via Conventional Communication Channels and the Possibility of Multiplexing in the Implementation of 3D Hyperspectral Images", *Photonics, MDPI*, vol. 8, pp. 448, 2021.
100. A. Manolova, K. Tonchev, V. Poulkov, S. Dixir, and P. Lindgren, "Context-Aware Holographic Communication Based on Semantic Knowledge Extraction", *Wireless Pers Commun., Springer*, vol. 120, pp. 2307–2319, 2021.
101. A. Shahraki, M. Abbasi, Md. J. Piran, and A. Taherkordi, "A Comprehensive Survey on 6G Networks: Applications, Core Services, Enabling Technologies, and Future Challenges, 2022. [Online]. Available: arXiv:2101.12475v2
102. P. K. Padhi, and F. Charrua-Santos, "6G Enabled Tactile Internet and Cognitive Internet of Healthcare Everything: Towards a Theoretical Framework", *Appl. Syst., Innov., MDPI*, vol. 4, pp. 66, 2021.
103. M. Kalfa, M. Gok, A. Atalik, B. Tegin, T. M. Duman, and O. Arikan, "Towards goal-oriented semantic signal processing: Applications and future challenges", *Digital Signal Processing, Elsevier*, vol. 119, pp. 103-134, 2021.
104. Q. Lan, D. Wen, Z. Zhang, Q. Zeng, X. Chen, P. Popovski, and K. Huang, "What Is Semantic Communication? A View on Conveying Meaning in the Era of Machine Intelligence", *Journal of Communications and Information Networks*, vol. 6, no. 4, pp. 336-371, 2021.